\documentclass[preprintnumbers,amsmath,11pt,amssymb,floatfix,superscriptaddress,nofootinbib]{article}
\usepackage{array}
\usepackage{relsize}
\usepackage{cite}
\usepackage{amsmath}
\usepackage{amssymb}
\usepackage{graphicx}
\def\mytitle#1{\setcounter{equation}{0}
\setcounter{footnote}{0}
\begin{flushleft}\Large\textbf{#1}\end{flushleft}
\vspace{0.25cm}}
\def\myname#1{\leftline{{\large #1}}\vspace{-0.13cm}}
\def\myplace#1#2{\small\begin{flushleft}\textit{#1}\\
\texttt{#2}\end{flushleft}}

\def\myclassification#1{\small\noindent
PACS:
       #1\vspace{0.5cm}}
\usepackage{graphicx}
\begin{document}

\mytitle{Threshold Drop in Accretion Density if Dark Energy is Accreting onto a Supermassive Black Hole}

\myname{Ritabrata Biswas\footnote{biswas.ritabrata@gmail.com} and Sandip Dutta \footnote{duttasandip.mathematics@gmail.com}}
\myplace{Department of Mathematics, The University of Burdwan, Burdwan -713104, West Bengal, India.}{} 
 
\begin{abstract}
Recent studies of galactic cores tell us that supermassive black holes are hosted at each of these cores. We got some evidences even. Besides, dark energy is expected to be distributed all over in our universe. Dark matter halo, on the other hand, could be found around the galactic regions. Though the natures of spans of them are not clearly measured. Galactic structures are supposed to be formed out of dark matter clustering. Some examples of supermassive black holes in the central regions of high redshift galaxies say that the concerned supermassive black holes have completed their constructions in a time less than it generally should be. To justify such discrepancies, we are forced to model about existences of black hole mimickers and exotic phenomena acting near the supermassive black holes. Motivated by these we study the natures of exotic matters, especially dark energy near the black holes. We choose modified Chaplygin gas as dark energy candidate. Again, the descriptions of gravitational waves or the attenuations of them when they are tunnelling through cosmological distances help us to measure the shear viscosity of the medium through which the waves have been travelled. Delayed decaying models of dark matters also suggest that dark energy and viscosity may come up as a byproduct of such decays or interactions. We consider the viscous nature of the medium, i.e., the dark energy. To do so, we choose an alpha-disc model as proposed by Shakura and Sunyaev. We study the variations of densities through accretion and wind branches for a different amount of viscosity regulated by the Shakura-Sunyaev's alpha parameter, spin parameter and different properties of accreting fluids, viz, the properties of adiabatic fluid and modified Chaplygin gas. We compare these results with each other and some existing density profiles drawn from observational data-based simulations. We follow that our result supports the data observed till date. Specifically, we see the wind to get stronger for dark energy as accreting agent. Besides, we see the accretion to have a threshold drop if the viscosity is chosen along with the repulsive effects of dark energy.

\end{abstract}

\myclassification{ 95.30.Sf,95.36.+x, 95.35.+d, 98.80.Cq,  98.80.-k }

\section{Introduction}
One of the very important predictions of Einstein's General Relativity (GR hereafter) is about the existence of a particular type of compact objects, the name of which is popularly coined as ``Black Holes" (BHs hereafter). BHs are nothing but such a dense object, the escape velocity from the surface of which is more than the speed of light in vacuum, $c$ (for simplest example of such objects known as Schwarzschild BH, calculated by Schwarzschild in 1916, $v_{escape}=\sqrt{\frac{2GM}{R^2}}\geq c$, i.e., $R\leq \frac{2G M}{c^2}$, $G$ is Newton's gravitational constant, $M$ is the mass of the central gravitating object, $R$ is the radius of the object). There exist many indirect proofs of the existences of BHs drawn from the peculiar behaviours and high energy physics of the infalling matters flowing towards the steep gravitational well around the concerned BH. With the help of observations of the ``Event Horizon Telescope", we are able to figure out the first-ever image of a BH's event horizon \cite{Eventhorizontelescope}. To state more correctly, what has actually been observed is the BH's shadow (a combination of event horizon and gravitational lensing of photons) not the event horizon itself. In any case, allow us to bring to readers' attentions on the work \cite{mail_gyan} where the authors have studied the possibility that the object observed might not be a Kerr BH but a more exotic object such as a superspinar, i.e., a BH spinning above the Kerr limit, which however would not necessarily leads to a naked singularity since quantum gravity effects could conspire to ``cover" the singularity (in fact it is claimed that such objects arise quite naturally in string theory, and some have even claimed that a detection of superspinning BHs could be a ``proof" of string theory). 

This particular supermassive BH is located at the central region of the elliptic galaxy  M87. The mass of this central BH, span of the accretion disc, the angular speed of the disc and accretion rate onto the BHare found to be roughly equal to $\left(6.5\pm 0.2_{stat}\pm 0.7_{phys}\right)\times 10^9M_{\odot},~~2.5\times 10^4 AU$, of the order $10^3kms^{-1}$ and $0.1M_{\odot}yr^{-1}$ respectively. Besides, recent gravitational wave detections\cite{Gravitational wave} of BH pair mergers can be treated to be proofs of supermassive BHs as well. 

Galactic structures through all over the universe possess two common features: the presence of a supermassive BH (SMBH hereafter) at the center and dark matter(DM) embeddings and surroundings all over the galaxy. The causes of these two phenomena, however, are not entirely clear to us. We are able to observe  SMBH of mass $\sim 800$ million $M_{\odot}$ even at redshift $z=7.54$ which is to be formed within a span less than a billion years only \cite{Eduardo Banados}. To justify such existences, alternative models of SMBHs are proposed. In the background of classical GR, different models for extended objects are considered, eg. gravastars and boson stars. Besides, existences of exotic cases like naked singularity etc are theoretically used to mimic SMBHs. Motivations of these concepts were to consider only the gravitational effects of the BH alternatives and to distinguish between the observational evidences drawn from the BHs and the BH mimickers. Authors of the reference \cite{K BoshkayevandD Malafarina} have shown, for Milky way, these two components' effects can not be distinguished in the inward region of a sphere persuing radius $\sim 100 AU$ distance from the center. If considerable amounts of pulsars are present in the vicinity of a SMBH, then only we might be able to distinguish between the BH shadow and the evidences of BH mimickers present near the BH. These researches strongly establish the presence of exotic matter/energy near the core regions of galaxies. The variation of densities for the model considered in the reference \cite{K BoshkayevandD Malafarina} is shown in figure (0). The density falls abruptly as we move far from the central region of the galaxy.  
\begin{figure}[h!]
\centering
~~~~~~~Fig $0$~~~~~~~\\
\includegraphics*[scale=.7]{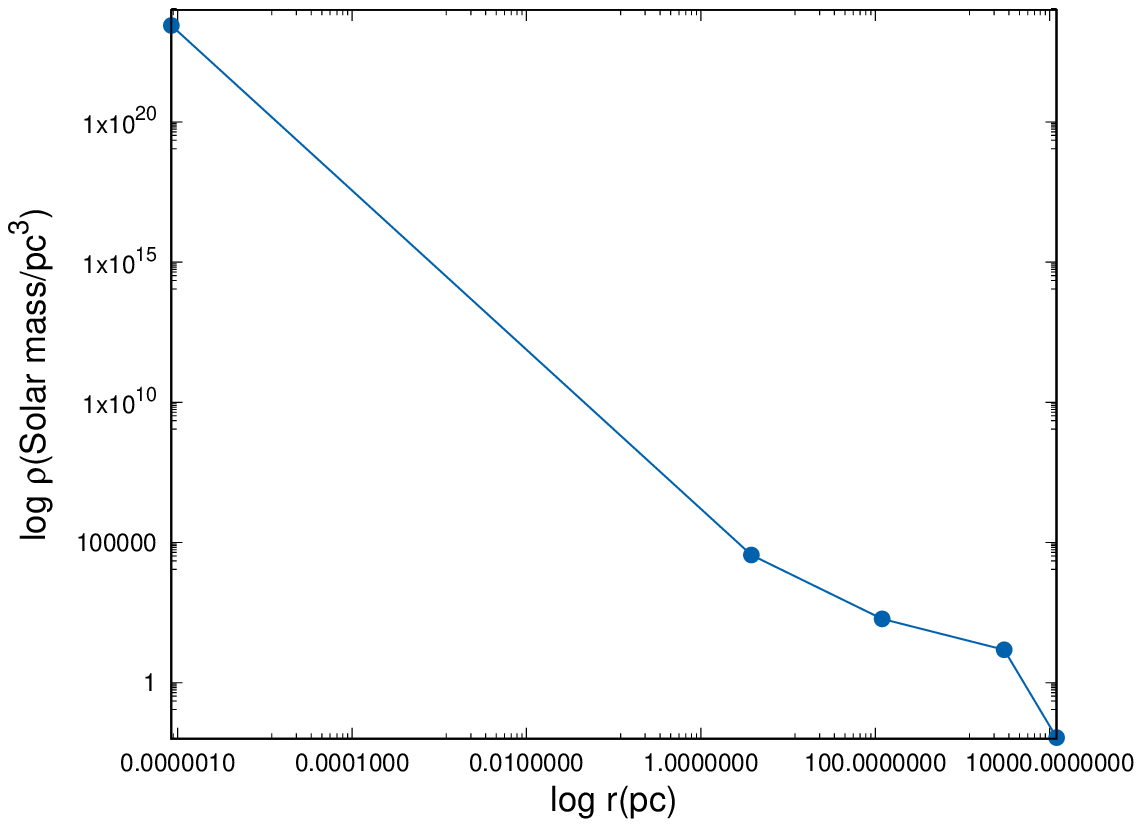}~~\\
\it{Figure 0 is variation of log of accretion density (in $\frac{M_{\odot}}{Pc^3}$ unit) vs log of radial distance (in $Pc$ unit) curve drawn from the data generated by the model considered in the reference \cite{K BoshkayevandD Malafarina}. We notice extremely high density near the core. As we go to outer regions, density falls abruptly.}
\end{figure}

``Dark energy" (DE hereafter) is a nomenclature to a hypothetical fluid which permeates all over the cosmos, exerting negative pressure to accelerate the late time expansion of the universe\cite{Peebles}. Interactive DM-DE models also support that if DM clusters are found in galactic centers, DE will grow from delayed decay of the DM. Along with this, the bulk viscosity is also supposed to grow in these regions \cite{PRLinteraction, PRDinteraction}.

Accretion flow of interstellar plasmas surrounding a relativistic star, i.e., a Schwarzschild singularity was first so ever studied by Michel \cite{Michel}. Before Michel, Bondi had worked on accretion properties of more or less Newtonian type and onto a nonrelativistic celestial object. Bondi did not consider the transonic properties of accretion disc either \cite{Bondi}. Michel has tried to plot the radial inward velocity($v_x$) of accretion flow for adiabatic fluids and has derived the energy flux of the flow. He has assumed the total internal energy of the infalling matter to get radiated out just before the infall. In Michel's work, the role of viscosity was neglected.

Shakura and Sunyaev \cite{ShakuraSunyaev1973} tried to estimate the quantity of infalling matter and to make simpler modelling of the angular momentum transport which builds up the thin disc-like structure. They have predicted that the efficiency of the magnetic field etc grown in a disc can not exceed the thermal energy of the matter which again can be quantified through the speed of sound ($c_s$) in the said accreting fluid. Compared to the scale of radial distance from the center, the turbulence is homogeneous and isotropic. Calculations on molecular viscosity, laminar flow properties and magnetic field-angular momentum transport connection led them to the replacement of tangential stresses by the simplified term: $-{\cal W}_{r\phi}\sim \alpha_{SS}\rho c_s^2$, where $\alpha_{SS}$ is the well known Shakura Sunyaev parameter and $\rho$ is the density of the accreting fluid. Again, Michel or Shakura-Sunyaev neglected the GR effects on the curvature near BH except for the truncation of the thin disc inside the innermost stable orbit. A complete GR view of accretion, for the first time, was established by Novikov and Thorne \cite{NT}.

After we have done with the simplifications of the complex natures of accretion disc viscosity, the very next problem to get solved was to simplify the highly nonlinear GR equations pointed out by Novikov and Thorne. To consider a stationary flow ($\frac{\partial }{\partial t}\equiv 0$), i.e., to simplify the analysis we must replace the relativistic attracting force offered by the central BH with a pseudo Newtonian force. We also consider an axisymmetric system with coordinates $r$, $\phi$ and $z$ with $\frac{\partial }{\partial \phi}\equiv 0$. Besides, Shakura and Sunyaev's work was for thin accretion disc which must not sustain anymore if the accretion rate is enough high. Paczynski and Witta \cite{Paczynsky and Witta} have taken a way out where flat disc was considered, inner edge of which is extended somewhere between marginally bound circular orbit and last stable circular orbit controlled under a pseudo Newtonian potential, $\psi=-\frac{GM}{r-r_g}$, $r_g=\frac{2GM}{c^2}$, $r$ is the radial distance from the center of the gravitating object. Later, this potential which was constructed for Schwarzschild metric was generalized for Kerr metric by B. Mukhopadhyay  as \cite{pseudonewtonian}:  
\begin{equation}
{\cal F}_g\left({\cal X}, j\right)= \frac{\left({\cal X}^2-2j\sqrt{{\cal X}}+ j^2\right)^2}{{\cal X}^3\lbrace \sqrt{{\cal X}} \left({\cal X}-2\right)+j\rbrace ^2}~~,
\end{equation}
where ${\cal X}=\frac{r}{r_g}$, $j$ is dimensionless specific angular velocity for rotating BH.
As the rotational parameter, $j$ is incorporated in the pseudo-Newtonian potential, we can find different accretion studies \cite{Mukhopadhyay2003} which analyzed the effects of the spin parameter on the accretion disc. Viscous accretion study with this potential is constructed by the same group of researchers \cite{viscocitybmukherjee1} and analysis of sonic point, density, temperature and abundance of different light elements in the disc and the outflow are studied.

Besides, the scenario of ongoing studies of adiabatic accretion properties around BHs, Babichev E. et al have started the study \cite{Babichev2004 DE accretion PRL} of DE accretion. There are several DE candidates which have been proposed and worked with to support to late-time cosmic acceleration. Among them, Modified Chaplygin Gas (MCG hereafter) is a particular one. Variations of its characterizing parameters turn MCG to mimic radiation era to phantom epoch \cite{Bento, Eos1, Eos2}. The Equation of State (EoS hereafter) of MCG is given by 
\begin{equation}\label{EoS}
p_{MCG}=\alpha_{MCG} \rho_{MCG} -\frac{\beta_{MCG}}{\rho_{MCG}^{n_{MCG}}}~~,
\end{equation}
where $\alpha_{MCG}$, $\beta_{MCG}$ and $n_{MCG}$ are MCG parameters, $\rho_{MCG}$ is the energy density of MCG. Like other DE candidates, different properties of accretion of MCG onto BHs have also been studied\cite{Biswasaccretion1, Biswasaccretion2, Jamil1}. In literature, we can find works on Chaplygin gas accretion. Biswas, R. et. al.  \cite{Biswasaccretion1} have speculated that wind becomes stronger for DE accretion and matter is thrown out with a speed equal to that of light at a finite distance from the BH. This implies that the repulsive power of DE makes the accretion phenomena weaker. Rotation of the central engine, in addition, makes this fainting out process more efficient. This work was followed by Biswas, R. \cite{Biswasaccretion2}. In this article, it was speculated that the nonviscous accretion's density profile is weaker if MCG accretion is considered. On the other hand, the wind branch's density is very high near to the BH. Recently, Dutta, S. and Biswas, R. \cite{Sandip1} have shown that, if the viscosity is encountered, the wind branch achieves speed equal to that of light at a distance which is lesser than the nonviscous case. Again BH spin enhances this tendency. In this present letter, we will study the variations of density in accretion and wind branches. The variations of density of accreting exotic matter have studied by different authors. Density variations in the accretion of DM halo is a well-opted topic as DM, unlike DE, is preferable to form clusters\cite{Deimer, Salvador-Sole}.

In this letter, we will consider viscous accretion of DE on SMBHs and study the density profiles of both accretion and wind flows. We have taken both the non-rotating and rotating BHs as the central engines. MCG is chosen as the DE model. Every result has been compared with corresponding adiabatic cases.

We wish to study the effects of viscosity, the negative pressure of DE and rotation on the accretion, precisely, on the density of the accretion and wind flow around SMBH. Study of such density profiles may reveal the natures of accretion when an exotic matter is intended to accrete in. 

The next part of this letter comprises of the mathematical formulation of viscous/nonviscous flow and adiabatic/MCG gas upon rotating/non-rotating BHs. After that, in section 3, we will study the variations of density graphically. In the end, we will briefly discuss the results and conclude. 
\section{Mathematical Formulation of Viscous Dark Energy Accretion Model}
To construct the mathematical model of the problem, we have taken the equation of continuity for cylindrical accretion model as 
\begin{equation}\label{equation of continuity}
\frac{d}{d{\cal X}} \left({\cal X} u \Sigma\right) = 0 ~~,
\end{equation}
where $\Sigma$ is vertically integrated density given by,
$
\Sigma = I_c \rho_e h\left({\cal X}\right),
$
$I_c$= constant (related to equation of state of fluid)= $1$ for simplicity, 
$\rho_e$=density of the accreting fluid at equitorial plane, 
$h\left({\cal X}\right)$= half thickness of the disc. $u=u_{\cal X}=\frac{v_{\cal X}}{c}$, $v_{\cal X}$ is the radially inward speed of accretion. 
Next three equations  will be the three components of Navier Stokes equation :
\\(a) Radial momentum balance equation:
\begin{equation}\label{radial momentum balance}
u\frac{du}{d{\cal X}} + \frac{1}{\rho} \frac{dp}{d{\cal X}}-\frac{\lambda^2}{{\cal X}^3}+{\cal F}_g\left( {\cal X}, j\right)=0,
\end{equation}
(b) Azimuthal momentum balance equation:
\begin{equation}\label{azimuthal momentum balance}
u \frac{d\lambda}{d{\cal X}} = \frac{1}{{\cal X} \Sigma} \frac{d}{d{\cal X}} \left[ {\cal X}^2 \alpha_{SS} \left(p+\rho u^2\right) h\left({\cal X}\right) \right],
\end{equation}
where $\lambda$ is specific angular momentum.

Assuming the vertical equilibrium from the vertical component of Navier Stokes equation we get the expression for $h({\cal X})$ as
\begin{equation}\label{halfthikness}
h\left({\cal X}\right) = c_s \sqrt{\frac{{\cal X}}{{\cal F}_g}}.
\end{equation}
Last four equations together will complete the present model for three dependent variables $u$, $c_s$ and $\lambda$. Differentiating equation (\ref{EoS}) with respect to $\rho$ we have,
\begin{equation}\label{Eoq2}
c_s^2 = \frac{\partial p_{MCG}}{\partial \rho_{MCG}} = \alpha_{MCG} + \frac{\beta_{MCG} n_{MCG}}{\rho_{MCG}^{n_{MCG}+1}}.
\end{equation}
On differentiation with ${\cal X}$ and using equation (\ref{Eoq2}) we obtain
$$
2c_s\frac{dc_s}{d{\cal X}} = -\left(n_{MCG}+1\right)\frac{\beta_{MCG}n_{MCG}}{\rho^{n_{MCG}+2}}\frac{d\rho}{d{\cal X}} =-\left(n_{MCG}+1\right)\left(c_s^2-\alpha_{MCG}\right)\times\frac{1}{\rho}\frac{d\rho}{d{\cal X}}
$$
which yields 
\begin{equation}\label{Eoq22}
\frac{1}{\rho}\frac{d\rho}{d{\cal X}}=-\frac{2c_s}{\left(n_{MCG}+1\right)\left(c_s^2-\alpha_{MCG}\right)}\frac{dc_s}{d{\cal X}}~~.
\end{equation}
Now, we will try to replace the pressure gradient over density term, i.e., the second term of the equation (\ref{radial momentum balance}), $\frac{1}{\rho}\frac{dp}{d{\cal X}}$, as
\begin{equation}\label{Eoq23}
\frac{1}{\rho}\frac{d\rho}{d{\cal X}}=\frac{1}{\rho}\frac{dp}{d{\rho}}\frac{d\rho}{d{\cal X}}=c_s^2\left(\frac{1}{\rho}\frac{d\rho}{d{\cal X}}\right)~~.
\end{equation}
Combining (\ref{Eoq22}) and (\ref{Eoq23}), and rearranging we have
$$\frac{1}{\rho_{MCG}} \frac{dp_{MCG}}{d{\cal X}}=  - \frac{2 c_s^3}{\left(n_{MCG}+1\right) \left( c_s^2 - \alpha_{MCG} \right)} \frac{d c_s}{d{\cal X}}=- \frac{2 c_s}{\left(n_{MCG}+1\right)} \frac{d c_s}{d{\cal X}}-\frac{\alpha}{\left(n_{MCG}+1\right)} \frac{d }{d{\cal X}}\left\{\ln \left(c_s^2-\alpha_{MCG}\right)\right\}$$
\begin{equation}\label{Eoq3}
 =-\frac{1}{n_{MCG}+1} \frac{d}{d{\cal X}} \left(c_s^2 \right) - \frac{\alpha_{MCG}}{n_{MCG}+1} \frac{d}{d{\cal X}} \lbrace ln \left( c_s^2 -\alpha_{MCG} \right) \rbrace.
\end{equation}
Now integrating the equation (\ref{equation of continuity}), we get the mass conservation equation
\begin{equation}\label{mass conservation}
\dot{{\cal M}} = \Theta \rho c_s \frac{{\cal X}^\frac{3}{2}}{{\cal F}_g^\frac{1}{2}} u,
\end{equation}
where $\Theta$ is geometrical constant.

Replacing the value of $\rho$ from equation (\ref{Eoq2}) in (\ref{mass conservation}) and differentiating the whole term we get a differential equation for $c_s$ as
\begin{equation}\label{differential equation for c}
\frac{d c_s}{d{\cal X}} = \left( \frac{3}{2{\cal X}} -\frac{1}{2{\cal F}_g} \frac{d{\cal F}_g}{d{\cal X}} + \frac{1}{u} \frac{du}{d{\cal X}} \right) \left\lbrace \frac{\left( n_{MCG}+1 \right) c_s \left( c_s^2 -\alpha_{MCG} \right)}{\left( 1-n_{MCG} \right) c_s^2 + \alpha_{MCG} \left( n_{MCG}+1 \right)} \right\rbrace 
\end{equation}
and from the equations (\ref{azimuthal momentum balance}) and (\ref{EoS}) we get the specific angular momentum gradient as
\begin{multline}\label{differential equation for lambda}
\frac{d \lambda}{d{\cal X}} = \frac{{\cal X} \alpha_{SS}}{u} \left[ \frac{1}{2} \left( \frac{5}{{\cal X}} - \frac{1}{{\cal F}_g} \frac{d{\cal F}_g}{d{\cal X}} \right) \left\lbrace \frac{\left(n_{MCG}+1 \right) \alpha_{MCG} - c_s^2}{n_{MCG}} + u^2 \right\rbrace \right.\\
\left. + 2 u \frac{du}{d{\cal X}} + \left\lbrace \left( \frac{\left(n_{MCG}+1 \right) \alpha_{MCG} - c_s^2}{n_{MCG}} + u^2 \right)  \frac{1}{c_s} -\left( c_s^2 +u^2 \right) \left( \frac{1}{n_{MCG}+1} \frac{2 c_s}{c_s^2 - \alpha_{MCG}} \right) \right\rbrace \frac{dc_s}{d{\cal X}} \right]. 
\end{multline}
Thus we get all the required values to replace $\frac{1}{\rho} \frac{dp}{d{\cal X}}$ in the equation (\ref{radial momentum balance}), hence we obtain the radial velocity gradient as 
\begin{equation}\label{differential equation for u}
\frac{du}{d{\cal X}} = \frac{\frac{\lambda^2}{{\cal X}^3}- {\cal F}_g\left( {\cal X} \right) + \left( \frac{3}{{\cal X}} - \frac{1}{{\cal F}_g} \frac{d{\cal F}_g}{d{\cal X}} \right) \frac{ c_s^4}{\lbrace \left( 1-n_{MCG} \right) c_s^2 + \alpha_{MCG} \left( n_{MCG}+1 \right)\rbrace}}{u - \frac{2 c_s^4}{u \lbrace \left( 1-n_{MCG} \right) c_s^2 + \alpha_{MCG} \left( n_{MCG}+1 \right)\rbrace}}.
\end{equation}
It is clear to observe that the denominator turns zero as $u=\frac{\sqrt{2}c_s^2}{\left\{(1-n_{MCG})c_s^2+\alpha_{MCG}(1+n_{MCG})\right\}^{\frac{1}{2}}}$ and this value lies in the interval $[0,~1]$ for some chosen ${\cal X}={\cal X}_c$(as $c_s$ lies in $[0,~1]$). Vanishing denominator signifies an unphysical flow. But as we consider the flow to be a continuous one, the numerator should simultaneously vanish at ${\cal X}={\cal X}_c$.

Applying the L'Hospital's rule  in equation (\ref{differential equation for u}) at the critical point (say ${\cal X}={\cal X}_c$) we get a quadratic equation of $\frac{du}{d{\cal X}}$ in the form,
\begin{equation}\label{quadratic}
{\cal A} \left( {\frac{du}{d{\cal X}}}\right)_{{\cal X}={\cal X}_c}^2 + {\cal B} \left( \frac{du}{d{\cal X}} \right)_{{\cal X}={\cal X}_c} + {\cal C} =0.
\end{equation}
Where \\
${\cal A}= 1 + \frac{1}{c_{sc}^2} - \frac{4 \left( n_{MCG}+1 \right) \left( c_{sc}^2 -\alpha_{MCG} \right)}{{\left\lbrace \left( 1-n_{MCG} \right) c_{sc}^2 +\alpha_{MCG} \left( n_{MCG}+1 \right) \right\rbrace}^2}$,\\
${\cal B}= \frac{4 \lambda u}{{\cal X}_c^3} + \frac{2 u_c c_{sc} \left( 1-n_{MCG} \right)}{ \left\lbrace \left( 1-n_{MCG} \right) c_{sc}^2 +\alpha_{MCG} \left( n_{MCG}+1 \right) \right\rbrace} - \frac{2 u_c \left( n_{MCG}+1 \right) \left( c_{sc}^2 -\alpha_{MCG} \right)}{{\left\lbrace \left( 1-n_{MCG} \right) c_{sc}^2 +\alpha_{MCG} \left( n_{MCG}+1 \right) \right\rbrace}^2} \left\{ \frac{3}{{\cal X}_c} - \frac{1}{{\cal F}_g} \left(\frac{d{\cal F}_g}{d{\cal X}}\right)_{{\cal X}={\cal X}_c} \right\} + \\
\left[ \left( c_{sc}^2 +u_c^2 \right) \left( \frac{1}{n_{MCG}+1} \frac{2 c_{sc}}{c_{sc}^2 - \alpha_{MCG}} \right)-\left\{ \frac{3}{{\cal X}_c} - \frac{1}{{\cal F}_g} \left(\frac{d{\cal F}_g}{d{\cal X}}\right)_{{\cal X}={\cal X}_c} \right\} \frac{4 u_c }{c_{sc}{\left\lbrace \left( 1-n_{MCG} \right) c_{sc}^2 +\alpha_{MCG} \left( n_{MCG}+1 \right) \right\rbrace}}-  \left( \frac{\left(n_{MCG}+1 \right) \alpha_{MCG} - c_{sc}^2}{n_{MCG}} + u^2 \right)  \frac{1}{c_{sc}} \right]\\ \times \left\lbrace \frac{ u_c \left( n_{MCG}+1 \right) \left( c_{sc}^2 -\alpha_{MCG} \right)}{2 c_{sc}^3} \right\rbrace $\\
and ${\cal C}= {\cal D}+{\cal E}+{\cal F}$.\\
The values of ${\cal D}$, ${\cal E}$ and ${\cal F}$ are,\\
${\cal D}= \left[ \left( c_{sc}^2 +u_c^2 \right) \left( \frac{1}{n_{MCG}+1} \frac{2 c_{sc}}{c_{sc}^2 - \alpha_{MCG}} \right)-\left\{ \frac{3}{{\cal X}_c} - \frac{1}{{\cal F}_g} \left(\frac{d{\cal F}_g}{d{\cal X}}\right)_{{\cal X}={\cal X}_c} \right\}\frac{4 u_c }{c_{sc}{\left\lbrace \left( 1-n_{MCG} \right) c_{sc}^2 +\alpha_{MCG} \left( n_{MCG}+1 \right) \right\rbrace}}\right.\\\left.-  \left( \frac{\left(n_{MCG}+1 \right) \alpha_{MCG} - c_{sc}^2}{n_{MCG}} + u^2 \right)  \frac{1}{c_{sc}} \right]
\left\{ \frac{3}{2{\cal X}_c} -\frac{1}{2{\cal F}_g} \left(\frac{d{\cal F}_g}{d{\cal X}} \right)_{{\cal X}={\cal X}_c} \right\} \left\lbrace \frac{\left( n_{MCG}+1 \right) c_{sc} \left( c_{sc}^2 -\alpha_{MCG} \right)}{\left( 1-n_{MCG} \right) c_{sc}^2 + \alpha_{MCG} \left( n_{MCG}+1 \right)} \right\rbrace $,\\
${\cal E}= \left\{ \frac{3}{{\cal X}_c} - \frac{1}{{\cal F}_g} \left(\frac{d{\cal F}_g}{d{\cal X}}\right)_{{\cal X}={\cal X}_c} \right\} \frac{u_c c_{sc} \left( 1-n_{MCG} \right)}{{\left\lbrace \left( 1-n_{MCG} \right) c_{sc}^2 +\alpha_{MCG} \left( n_{MCG}+1 \right) \right\rbrace}} + \left(\frac{d{\cal F}_g}{d{\cal X}}\right)_{{\cal X}={\cal X}_c} + \left\lbrace \frac{1}{{\cal F}_g^2} \left(\frac{d{\cal F}_g}{d{\cal X}} \right)^2_{{\cal X}={\cal X}_c} - \frac{1}{{\cal F}_g} \left(\frac{d^2 {\cal F}_g}{d{\cal X}^2}\right)_{{\cal X}={\cal X}_c} - \frac{3}{{\cal X}_c^2}\right\rbrace \frac{u_c^2}{2}$,\\
${\cal F}= \frac{\lambda \alpha_{ss}}{{\cal X}_c^2 u_c} \left\{ \frac{1}{{\cal F}_g} \left(\frac{d{\cal F}_g}{d{\cal X}}\right)_{{\cal X}={\cal X}_c} - \frac{5}{{\cal X}_c} \right\} \left\lbrace \frac{\left(n_{MCG}+1 \right) \alpha_{MCG} - c_{sc}^2}{n_{MCG}} + u_c^2 \right\rbrace  $.

Here $c_{sc}=c_s\left({\cal X}={\cal X}_c\right)$. We can obtain the accretion and wind fluid speeds by solving equations (\ref{differential equation for c}), (\ref{differential equation for lambda}) and ((\ref{differential equation for u})) starting from the critical point with proper initial conditions. Initial relation and hence the values of fluid speed and sound speed may be obtained by solving $N\left({\cal X}_c\right)=D\left({\cal X}_c\right)=0$ in the equation (\ref{differential equation for u}) due to the requirement of transonic smooth flow. $\lambda({\cal X}_c)=\lambda_c$ has been chosen artificially. 

We will try to find out density variations through the accretion and wind profiles. To do that we will use the concept of Eddington mass accretion rate. To keep an accretion process running, the considered central object should possess a luminosity up to a maximum limit beyond which the radiation pressure of that object will overcome gravity so that material outside the object will be swiped away from it, rather than falling inwards. This maximum luminosity is known as Eddington luminosity. To obtain this limit, we will balance the gravitational force with radiation force to obtain
$$\frac{GMm}{R^2}=F_{grav}=F_{rad}=P_{rad}\kappa m=\frac{L}{c}\frac{1}{4\pi R^2}\kappa m=\frac{L}{c}\frac{1}{4\pi R^2}\frac{\sigma_T}{m_p}m$$
where $\kappa$ is the opacity, $\sigma_T$ is Thompson scattering cross-section and $m_p$ is the mass of a proton. So
$L_{Edd}=\frac{4\pi G M c m_p}{\sigma_T}$. If $\epsilon$ fraction of the mass accretion rate is radiated out as energy, i.e., $L=\epsilon \dot{{\cal M}}c^2$ and we get Eddington mass accretion rate for a $10^7 M_{\odot}$ object as
$$\dot{{\cal M}}=\frac{4\pi G M m_p}{\epsilon c \sigma_T}=7.03592\times 10^{-16}\times \left(10^7 M_{\odot}\right) ~sec^{-1}~~.$$
From the equation (\ref{mass conservation}) we get the density in $gcm^{-3}$ as an equation of dimensionless quantities ${\cal X}$, $c_s$ and $u$ as 
\begin{equation}
\rho= \frac{2.285 \times 10^{-21} \times \sqrt{\frac{{\cal F}_g}{{\cal X}^3}}}{u c_s}~~~~.
\end{equation}
\section{Variations of Density in Accretion and Wind Branch : Solutions }
The DM distribution in galaxies and clusters of galaxies is well described by $\Lambda$CDM numerical n-body simulations. This theory also discussed cosmological evolution and the formation of astrophysical objects. These numerical simulations produce radial density profiles over a few decades in a radius near universal DM halo \cite{Navarro_2004, Stadel_2009}. Generally, Navarro-Frenk-White (NFW)\cite{Navarro_1997} profile supplies us a good fit to density profiles of equilibrium DM halo \cite{Diemand_2004, Diemand_2005a, Diemand_2005b, Diemand_2011, Ludlow_2011, Navarro_2010, Tasitsiomi_2004}. But some discrepancies do occur between simulations and observations. As an example, we can look up to the core problem \cite{Flores_1994} of N-body simulations which gives cuspy central density profiles \cite{Dubinski_1991} which can fit well by NFW profile, while observation provides core like centers in DM dominated systems like dwarf galaxies \cite{Pontzen_2014, Weinberg_2015}.

Usually, radial density profile is supposed to have a monotonically changing logarithmic slope such as NFW or Einasto profile \cite{Einasto_1965}. There are two different classes which exhibit non-monotonic changes in their density profile slopes \cite{young_2016}. The first class consists of the systems that have density oscillation defined through their distribution function $f(E)$, or differential energy distribution $N(E)$. Examples of this class are DARKexp profiles \cite{Hjorth_2010}, Polytropic \cite{Feron_2008, Medvedev_2001}, Isothermal sphere(IS) \cite{Binney_1987}, King Profile \cite{King_1966}. On the other hand, the members of the second class are the systems which are the product of dynamical evolution, like observed and simulated galaxies and pure DM halos. NFW profile and Einasto profile are the examples of this class.
\begin{figure}[h!]
\centering
~~~~~~~Fig $1.1.1.a$~~~~~~~\hspace{3 in}~~~~~~~~~Fig $1.1.1.b$
\includegraphics*[scale=0.3]{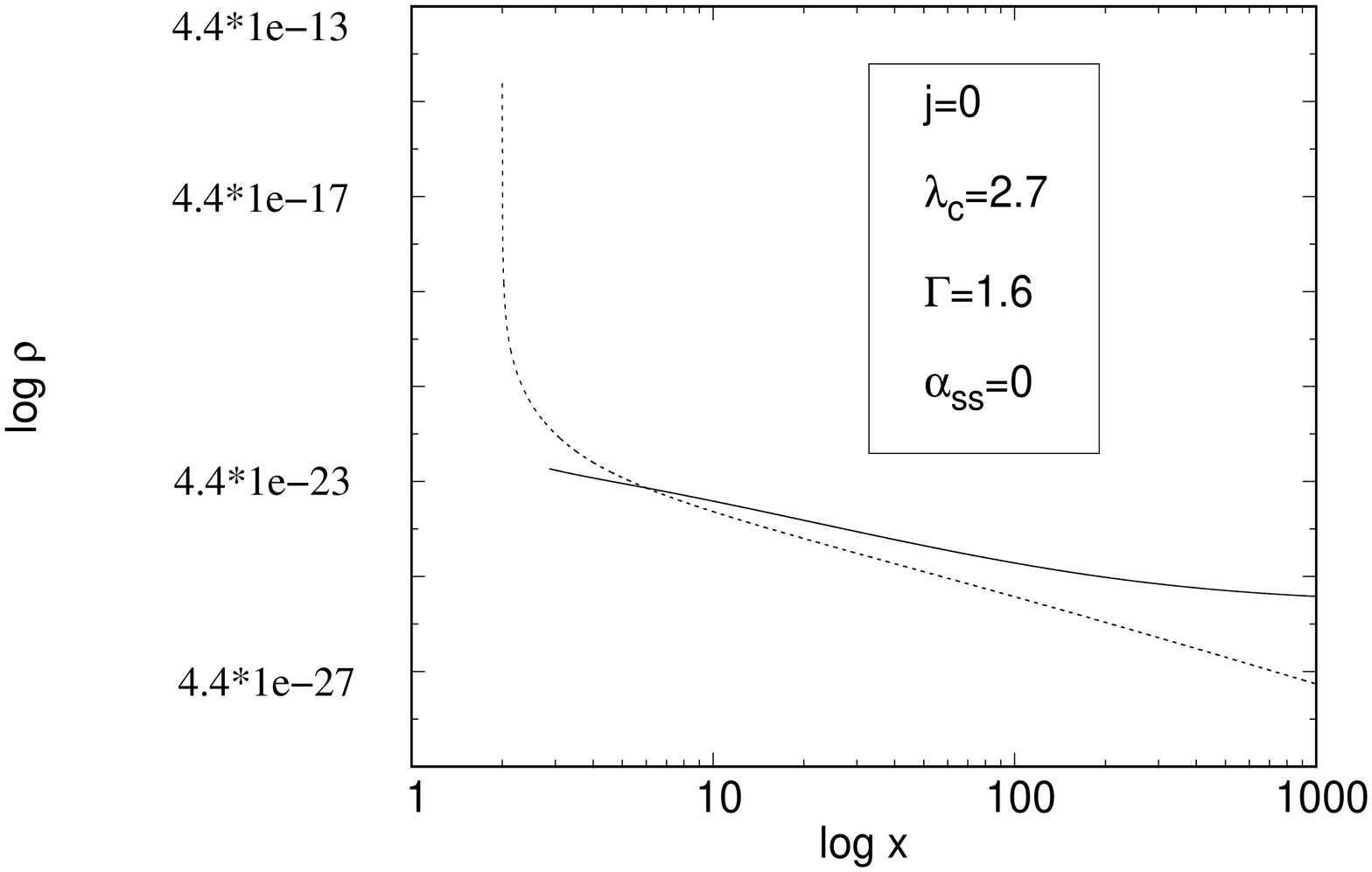}~~
\hspace*{0.2 in}
\includegraphics*[scale=0.3]{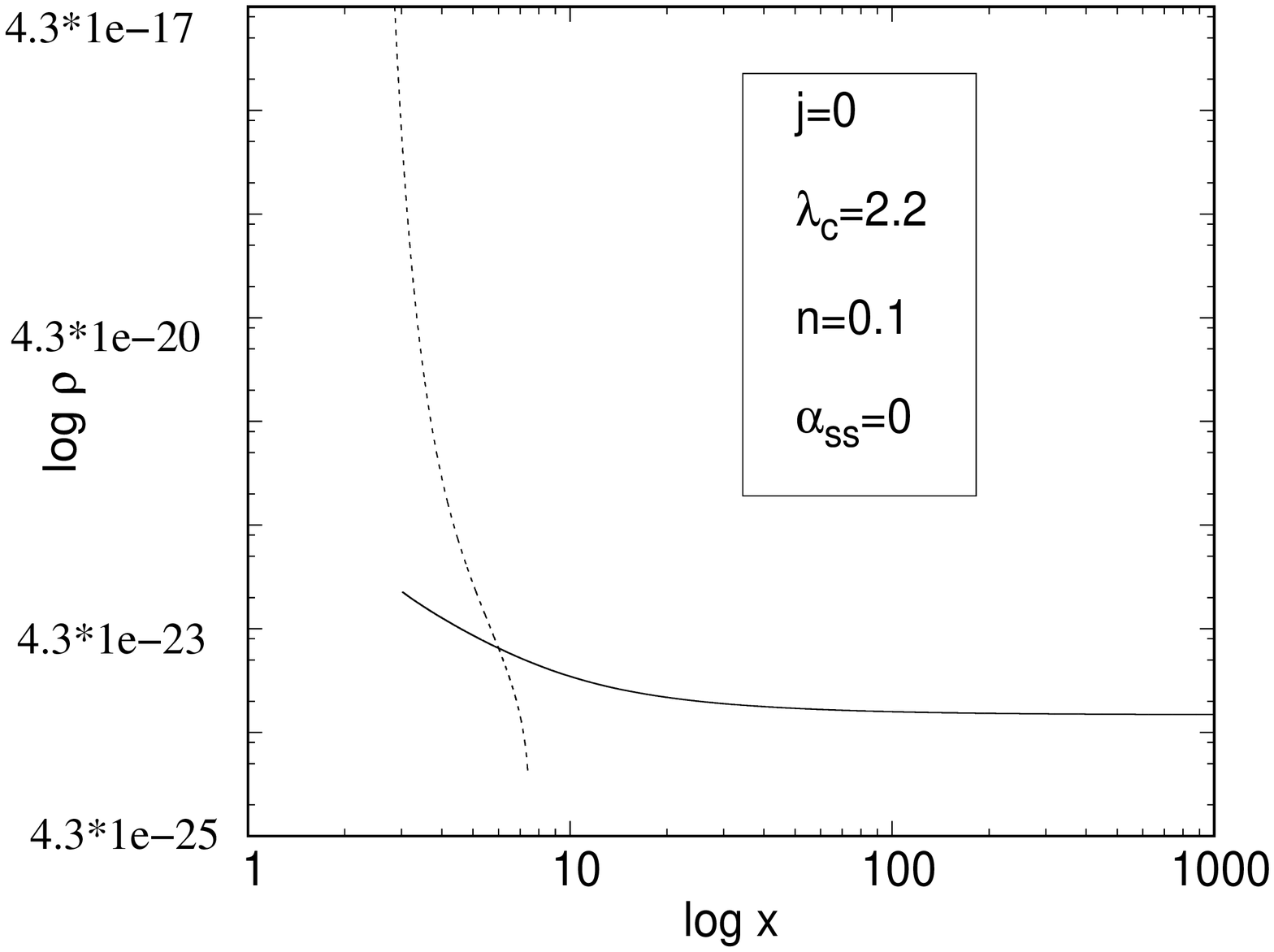}~~\\

~~~~~~~Fig $1.1.2.a$~~~~~~~\hspace{3 in}~~~~~~~~~Fig $1.1.2.b$
\includegraphics*[scale=0.3]{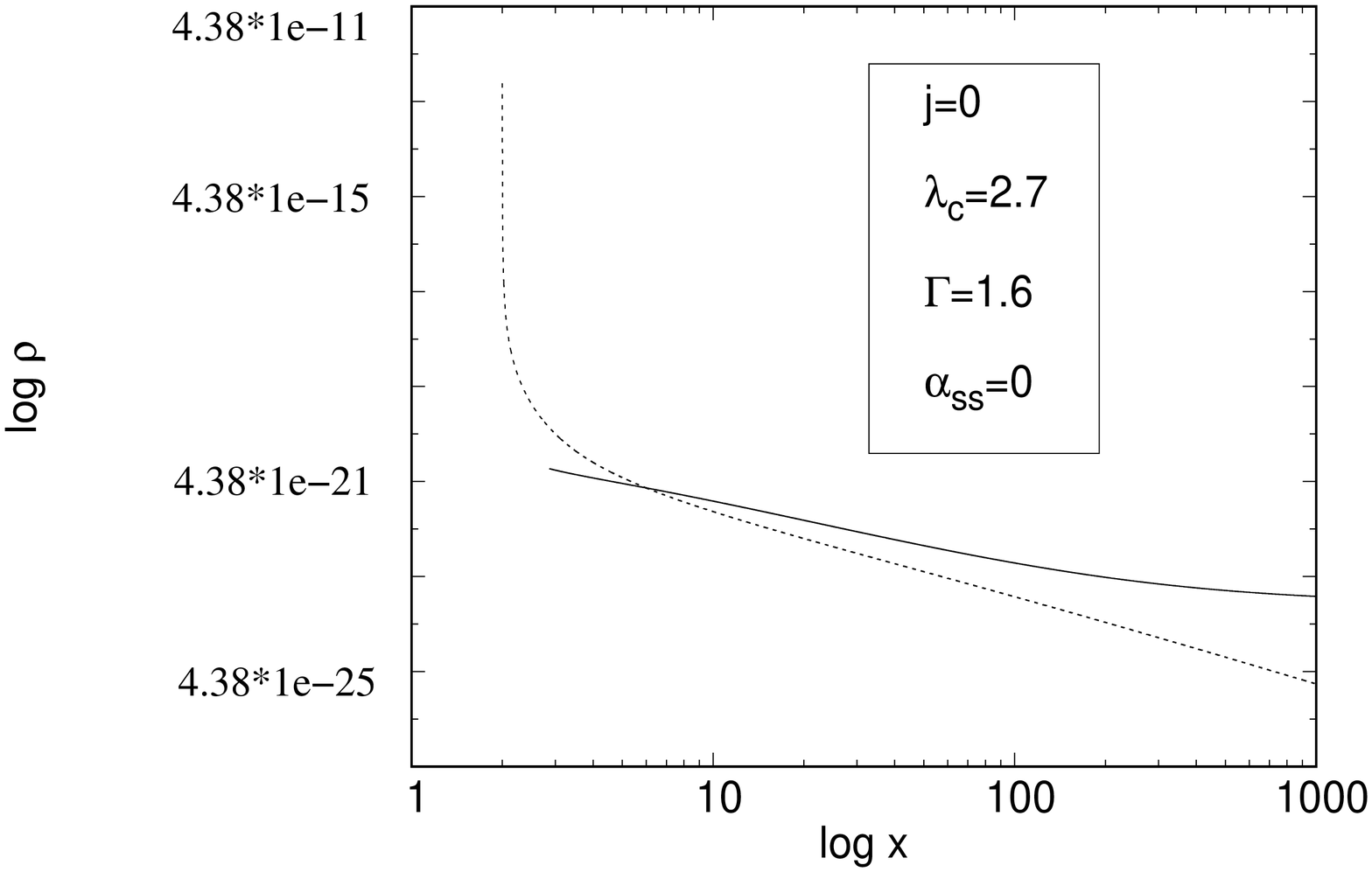}~~
\hspace*{0.2 in}
\includegraphics*[scale=0.3]{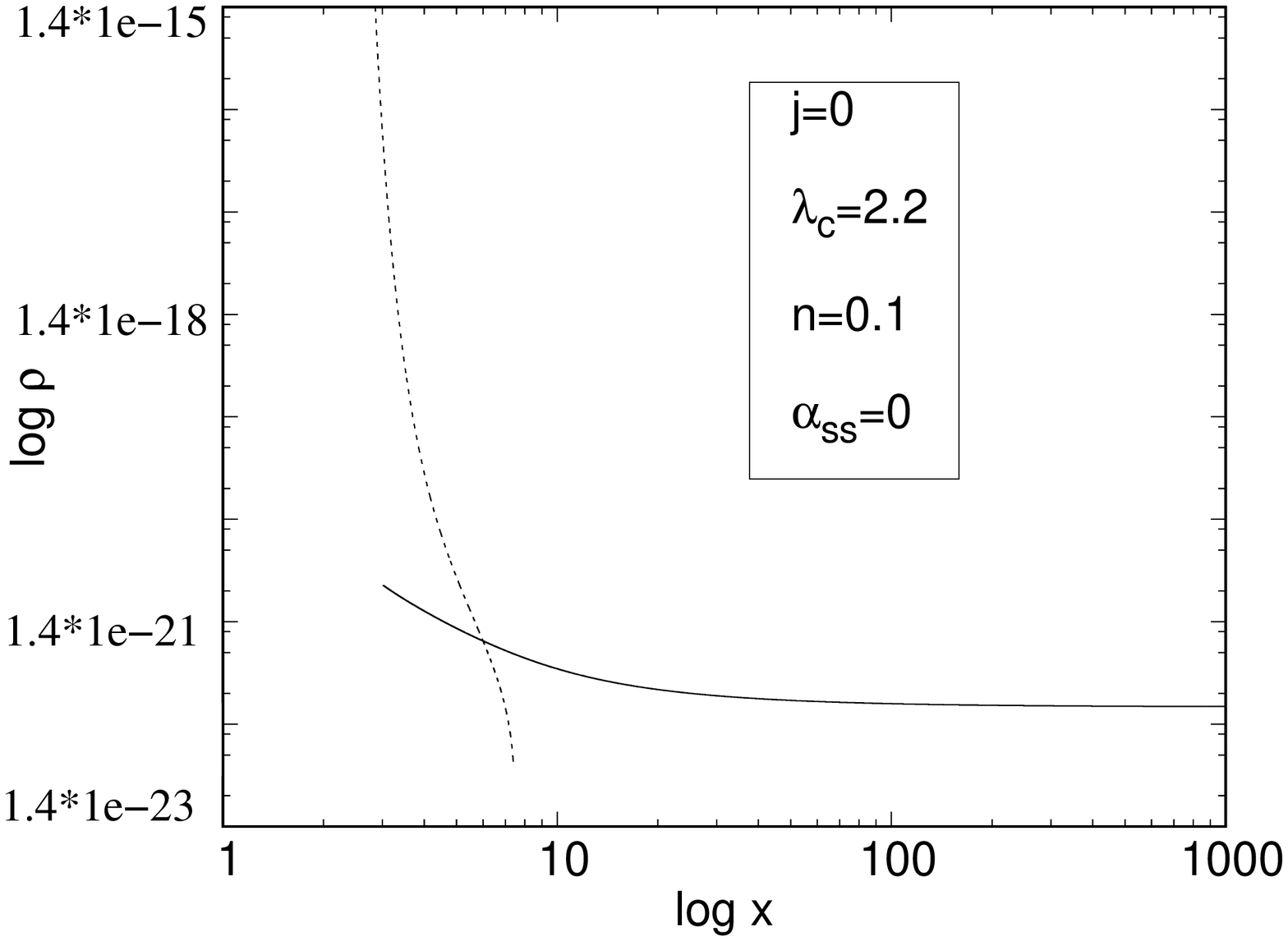}~~\\
\it{Figures $1.1.1.a$ and $1.1.2.a$ are plots of $log(\rho)$ vs $log(x)$ for nonviscous accretion disc flow around a non-rotating BH for adiabatic case using NFW and Einasto profile respevctively. Figures $1.1.1.b$ and $1.1.2.b$ are plots of $log(\rho)$ vs $log(x)$ for nonviscous accretion disc flow around a non-rotating BH for MCG using NFW and Einasto profile respectively. Accretion and wind curves are depicted by solid and dotted lines respectively.}
\end{figure}
Here, in our letter, we only consider NFW profile and Einasto profile as these two profiles give a good fit with the density profile of DM in the center of galaxy or cluster of galaxies. Navarro, Frenk and White used high-resolution simulations to study the formation of CDM halos with mass spanning about $4th$ orders of magnitude, ranging from dwarf galaxy halos to rich galaxy clusters. Their work showed that the equilibrium density profile of CDM halos of all masses can be accurately fitted by the simple formula \cite{Navarro_1997}.
\begin{equation}\label{Navarro}
\frac{\rho\left({\cal X} r_g\right)}{\rho_{crit}}= \frac{\delta_c}{ {\cal X} \left(1+ {\cal X}\right)^2},
\end{equation} 
where, $\rho_{crit}=\frac{3H^2}{8 \pi G}$ is the critical density and $\delta_c$ is a characteristic (dimensionless) density. The NFW simulation suggests that the density profile of an isolated equilibrium halo can be specified by giving two parameters, the mass of the halo and the characteristic density of the halo. In some particular hierarchical model, these parameters are related in such a way that the characteristic density is proportional to the mean cosmic density. Thus the characteristic density reflects the density of the universe at the collapse time of the objects, which merge to form the halo core.
Now using our model with the equation (\ref{Navarro}) we get the density $4.3\times 10^{-23} gcm^{-3}$ at a distance of six Schwarzschild radius distance from the center of the central engine. It is to be noted that the order of the density we get $(10^{-23})$ which is higher than the density of the universe (approximately of order $10^{-30}$).

Now we will study the Einasto profile. The formula for this profile is given by
\begin{equation}
\rho({\cal X} r_g)=\rho_s exp \left[ -b_n \left\lbrace ({\cal X})^{\frac{1}{n}}-1 \right\rbrace \right],
\end{equation}
where $\rho_s= \rho_{crit} \times \delta_c$, $n$= polytropic index, $b_n$ is a parameter depending on polytropic index. Einasto profile implies that the spherically averaged density profile of DM haloes are not universal but depends on the slope of DM power spectrum. The haloes, which are slowly growing through merger and accretion have steeper and more centrally concentrated density profiles with extended outer envelopes. Whereas the faster-growing haloes have shallower inner profiles and steep outer ones. 
We can obtain the density at $x=6$ with this formula both for adiabatic case and MCG. For adiabatic case, the density is $4.38\times 10^{-21} gcm^{-3}$ and for MCG it is $1.40\times 10^{-21} gcm^{-3}$.
\subsection{Graphical Interpretations of the Density Profiles}
We will plot density vs radial distance. Fig 1.1.1.a \& 1.1.2.a describes the density vs radial distance curves for adiabatic fluid without viscosity with NFW profiles and Einasto profile respectively. Besides, the concerned central engine is taken to be a non-rotating, i.e., a Schwarzschild one. The solid lines in the graphs are the plots of accretion density whereas the dashed curve is for the wind density. It is seen along with the accretion curve density increases as we go nearer to the BH. This is quite obvious due to the pressure felt in the steep gravitational field.

For wind, density at far is lower than that in the accretion curve and increases more rapidly than the accretion density and near the BH it is enough high. Near the BH, mass is piled up creating a throwback or outward shock for the infalling matter creating higher wind. This may raise the wind density near the BH. As we go far, outward throwing force decreases and density also decreases. 

Fig 1.1.1.b \& 1.1.2.b show the density vs radial distance curve for inviscid MCG accretion for non-rotating BH with NFW profiles and Einasto profile respectively. We see the wind is highly dense when it is near to BH. As it goes far from the BH, density falls very quickly. High density, even if for MCG produces less negative pressure. But as the density falls, negative pressure becomes high and the disc must have fainted. 
\begin{figure}[h!]
\centering
~~~~~~~Fig $1.2.1.a$~~~~~~~\hspace{3 in}~~~~~~~~~Fig $1.2.1.b$
\includegraphics*[scale=0.3]{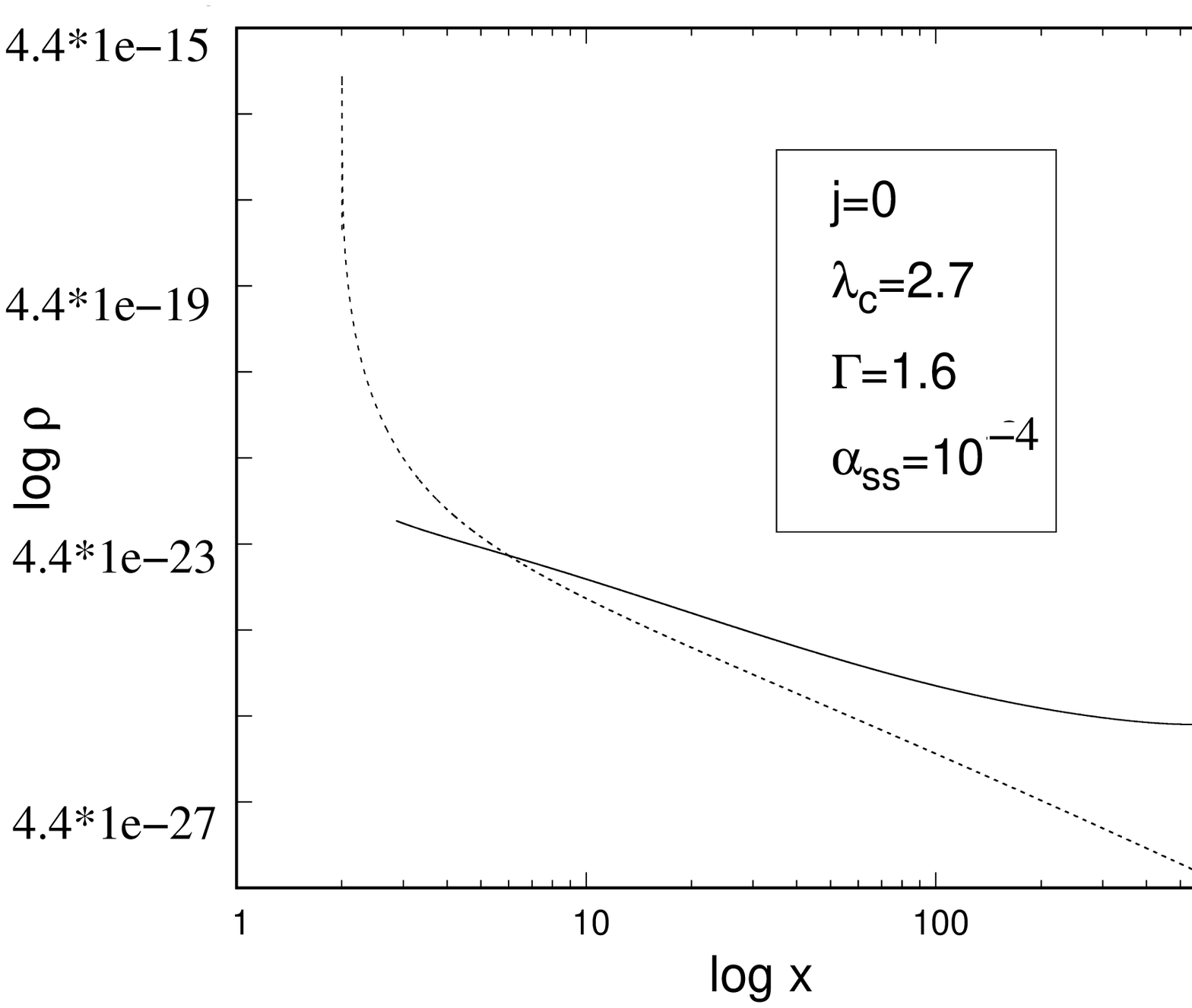}~~
\hspace*{0.2 in}
\includegraphics*[scale=0.3]{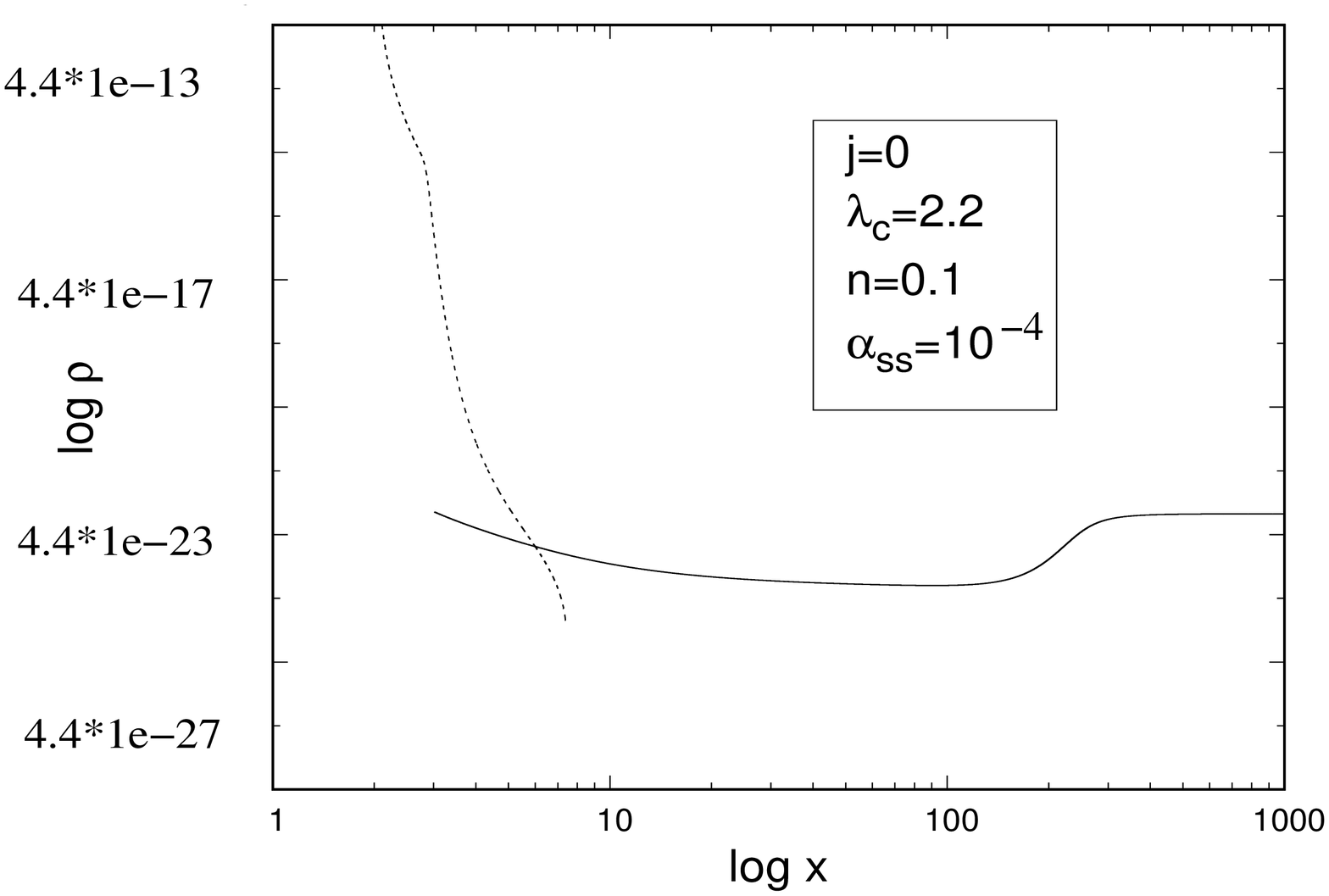}~~\\

~~~~~~~Fig $1.2.2.a$~~~~~~~\hspace{3 in}~~~~~~~~~Fig $1.2.2.b$
\includegraphics*[scale=0.3]{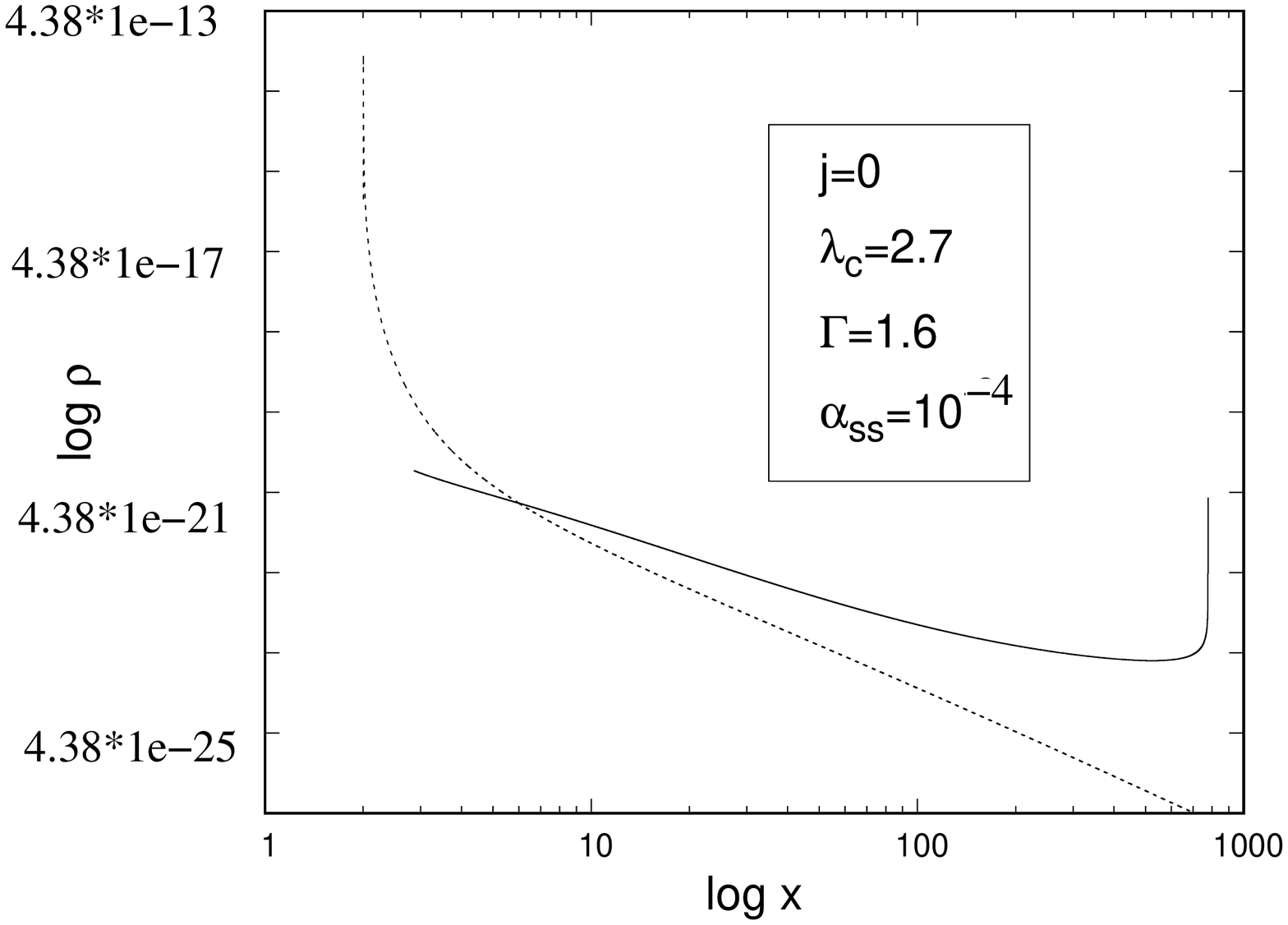}~~
\hspace*{0.2 in}
\includegraphics*[scale=0.3]{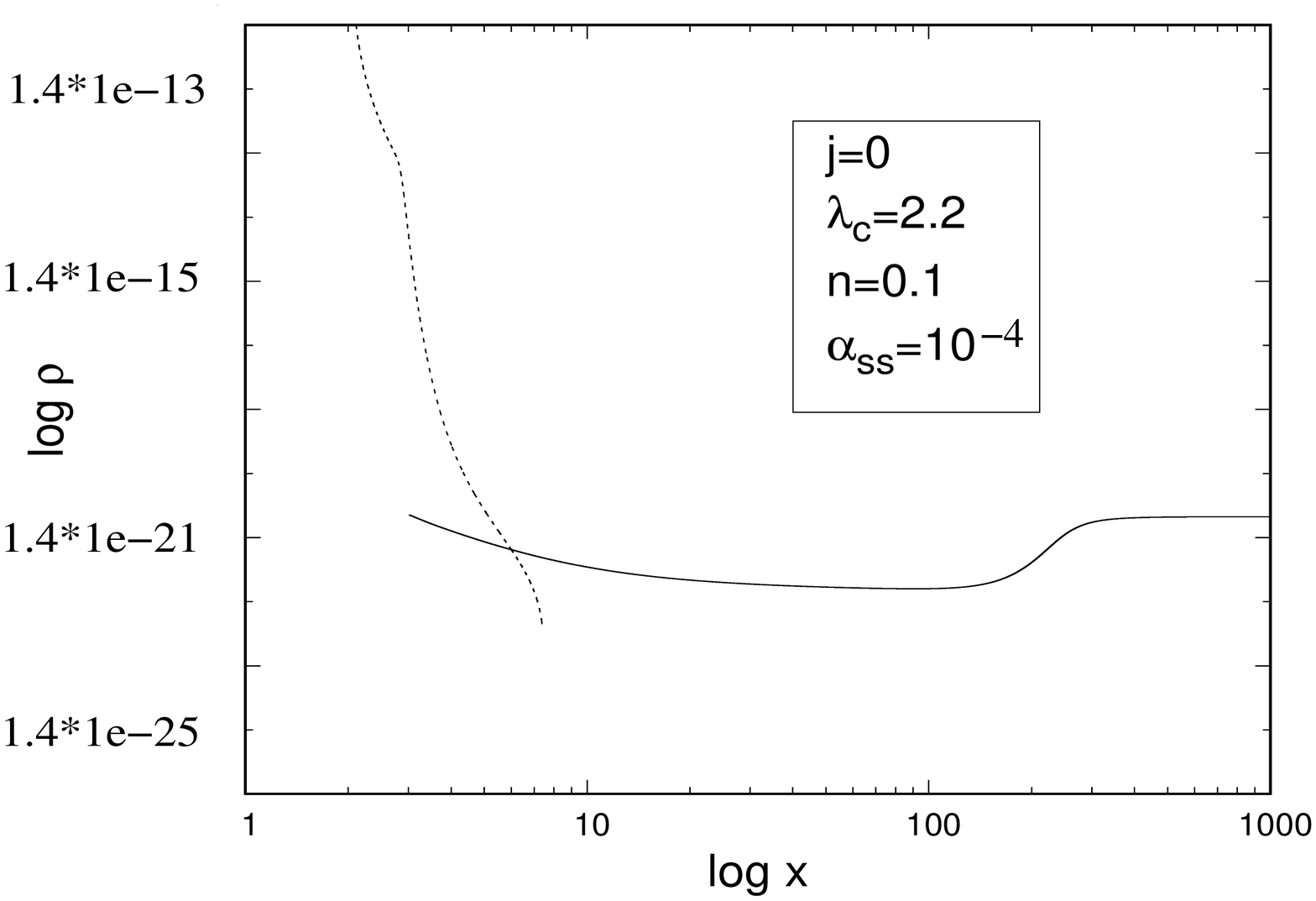}~~\\
\it{Figures $1.2.1.a$ and $1.2.2.a$ are plots of $log(\rho)$ vs $log(x)$ for nonviscous accretion disc flow around a non-rotating BH for adiabatic case using NFW and Einasto profile respevctively. Figures $1.2.1.b$ and $1.2.2.b$ are plots of $log(\rho)$ vs $log(x)$ for viscous accretion disc flow with viscosity $\alpha_{ss}=10^{-4}$, around a non-rotating BH for MCG using NFW and Einasto profile respectively. Accretion and wind curves are depicted by solid and dotted lines respectively.}
\end{figure}
\begin{figure}[h!]
\centering
~~~~~~~Fig $1.3.1.a$~~~~~~~\hspace{3 in}~~~~~~~~~Fig $1.3.1.b$
\includegraphics*[scale=0.3]{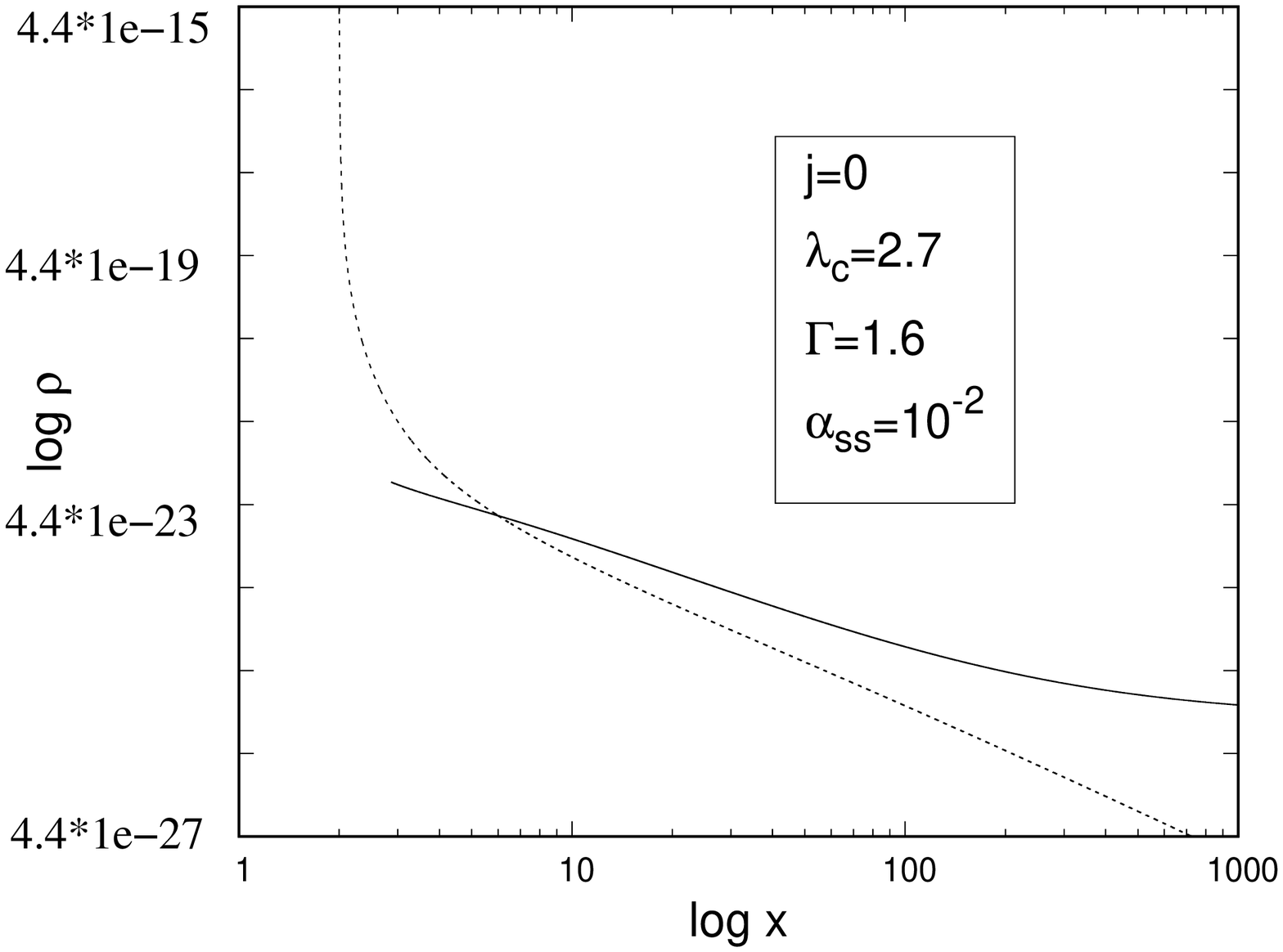}~~
\hspace*{0.2 in}
\includegraphics*[scale=0.3]{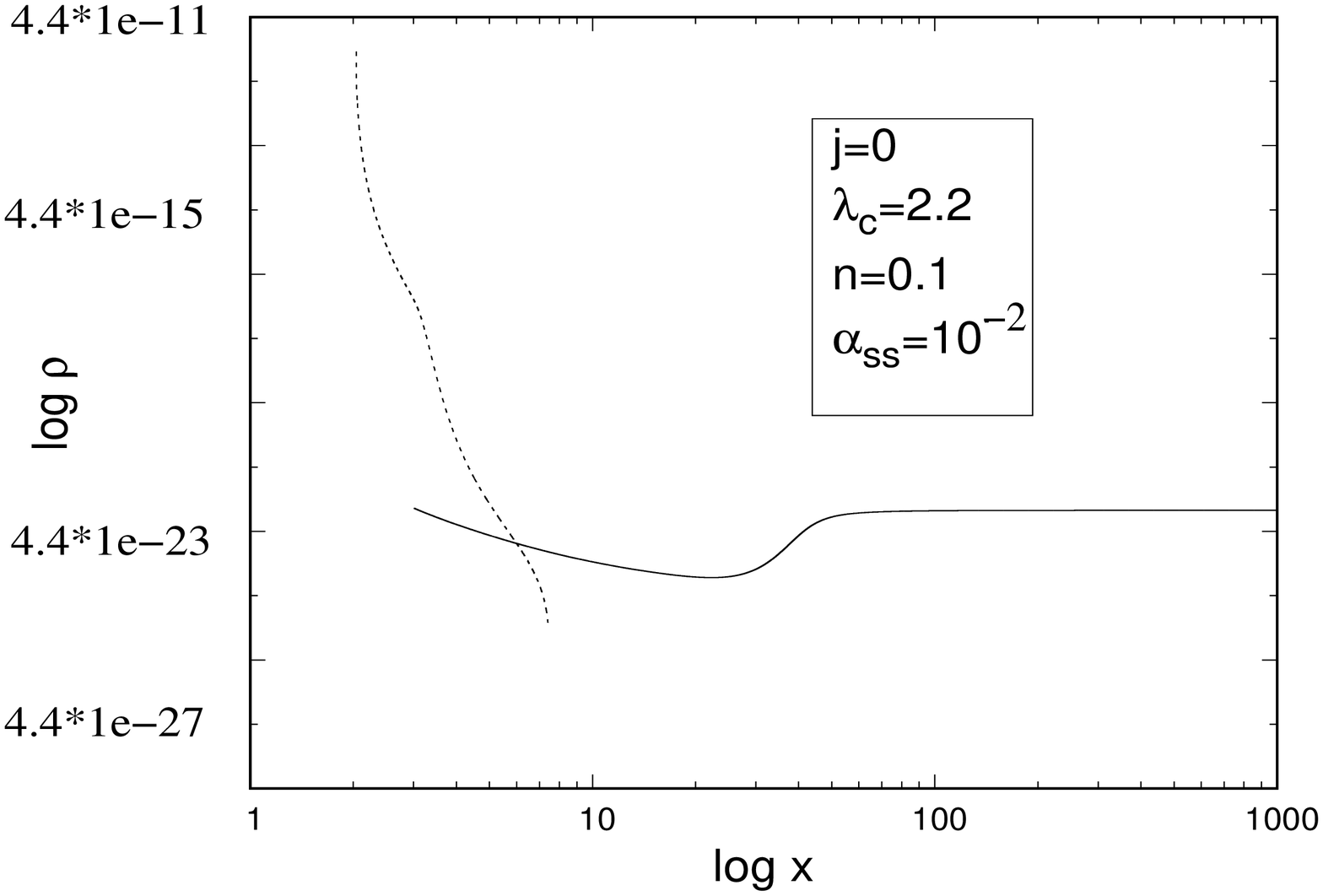}~~\\

~~~~~~~Fig $1.3.2.a$~~~~~~~\hspace{3 in}~~~~~~~~~Fig $1.3.2.b$
\includegraphics*[scale=0.3]{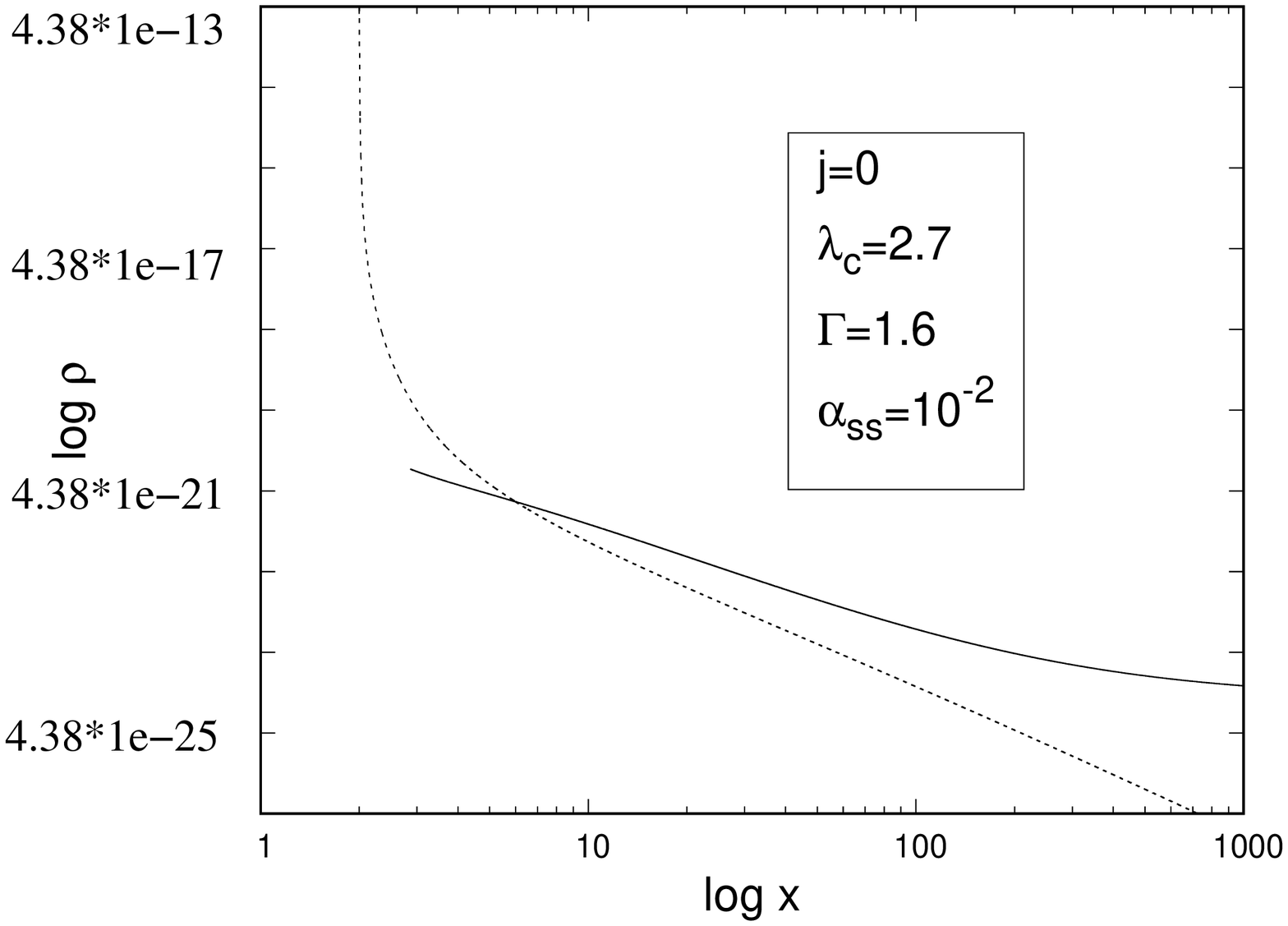}~~
\hspace*{0.2 in}
\includegraphics*[scale=0.3]{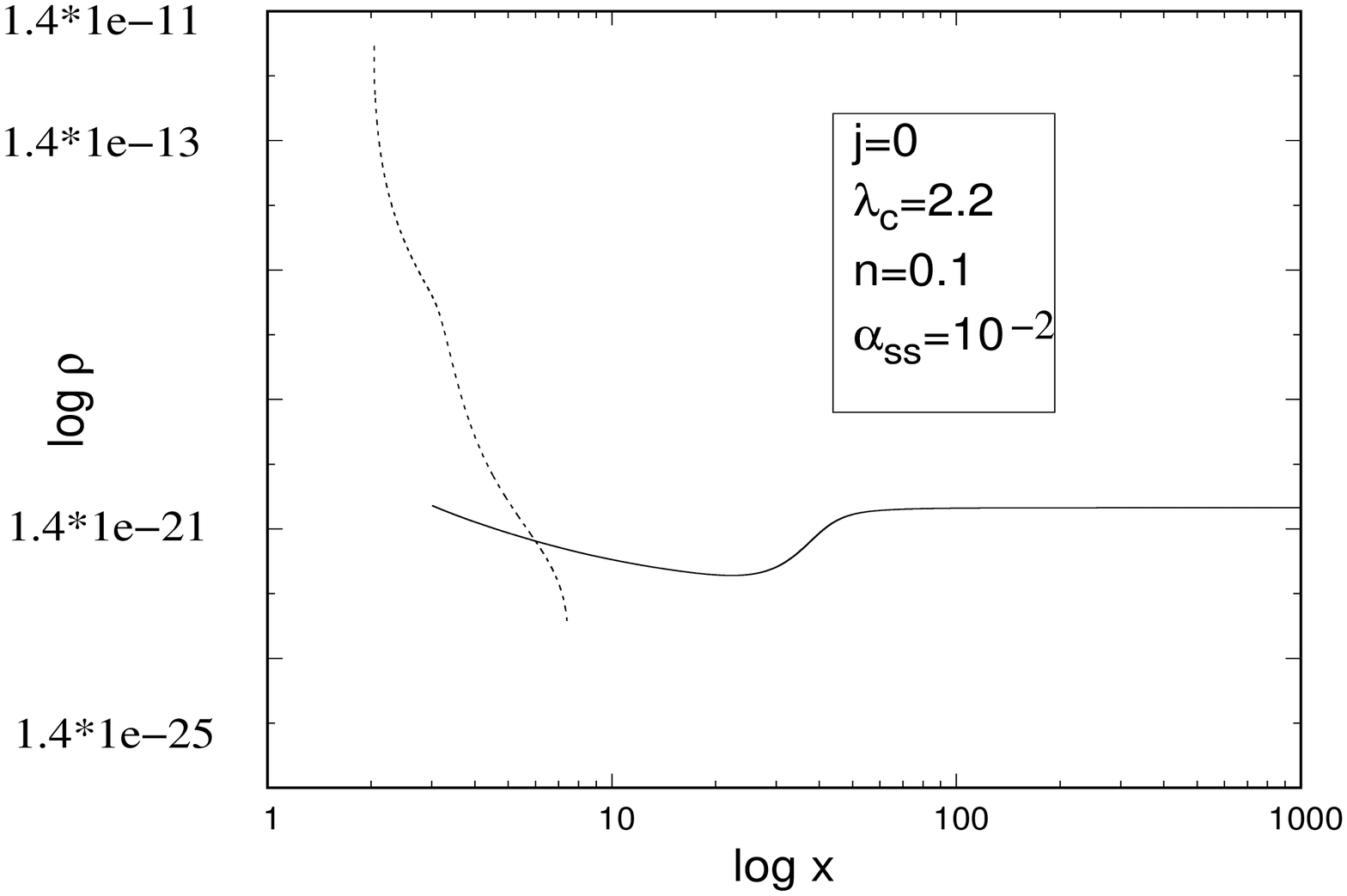}~~\\
\it{Figures $1.3.1.a$ and $1.3.2.a$ are plots of $log(\rho)$ vs $log(x)$ for nonviscous accretion disc flow around a non-rotating BH for adiabatic case using NFW and Einasto profile respevctively. Figures $1.3.1.b$ and $1.3.2.b$ are plots of $log(\rho)$ vs $log(x)$ for viscous accretion disc flow with viscosity $\alpha_{ss}=10^{-2}$, around a non-rotating BH for MCG using NFW and Einasto profile respectively. Accretion and wind curves are depicted by solid and dotted lines respectively.}
\end{figure}

Now we will plot the density vs radial distance curve for adiabatic flow with viscosity $\alpha_{ss}= 10^{-4}$ in fig 1.2.1.a \& 1.2.2.a for non-rotating BH and viscosity $\alpha_{ss}= 10^{-2}$ in Fig 1.3.1.a \& 1.3.2.a with NFW profiles and Einasto profile respectively. We see that the wind branch is less steeper, it has the highest value near the central engine and while we go far from the central engine the density decreasing spontaneously. We can also observe that the accretion branch changes its properties with increasing viscosity. At a certain distance, the density of accretion jumped over suddenly and the accretion disc no longer exists there.

In fig 1.3.1.b \& 1.3.2.b we plot the same graph with viscosity $\alpha_{ss}= 10^{-4}$ we show that the density of accretion has no notable change.

Fig 1.3.1.b \& 1.3.2.b show the density vs radial distance curve for MCG accretion with viscosity $\alpha_{ss}= 10^{-2}$ for non-rotating BH with NFW profiles and Einasto profile respectively. We see the wind is highly dense when it is near to BH. As it goes far from the BH, density falls very quickly. High density, even if for MCG produces less negative pressure. But as the density falls, negative pressure becomes high and the disc must have fainted. 
\begin{figure}[h!]
\centering
~~~~~~~Fig $2.1.1.a$~~~~~~~\hspace{3 in}~~~~~~~~~Fig $2.1.1.b$
\includegraphics*[scale=0.3]{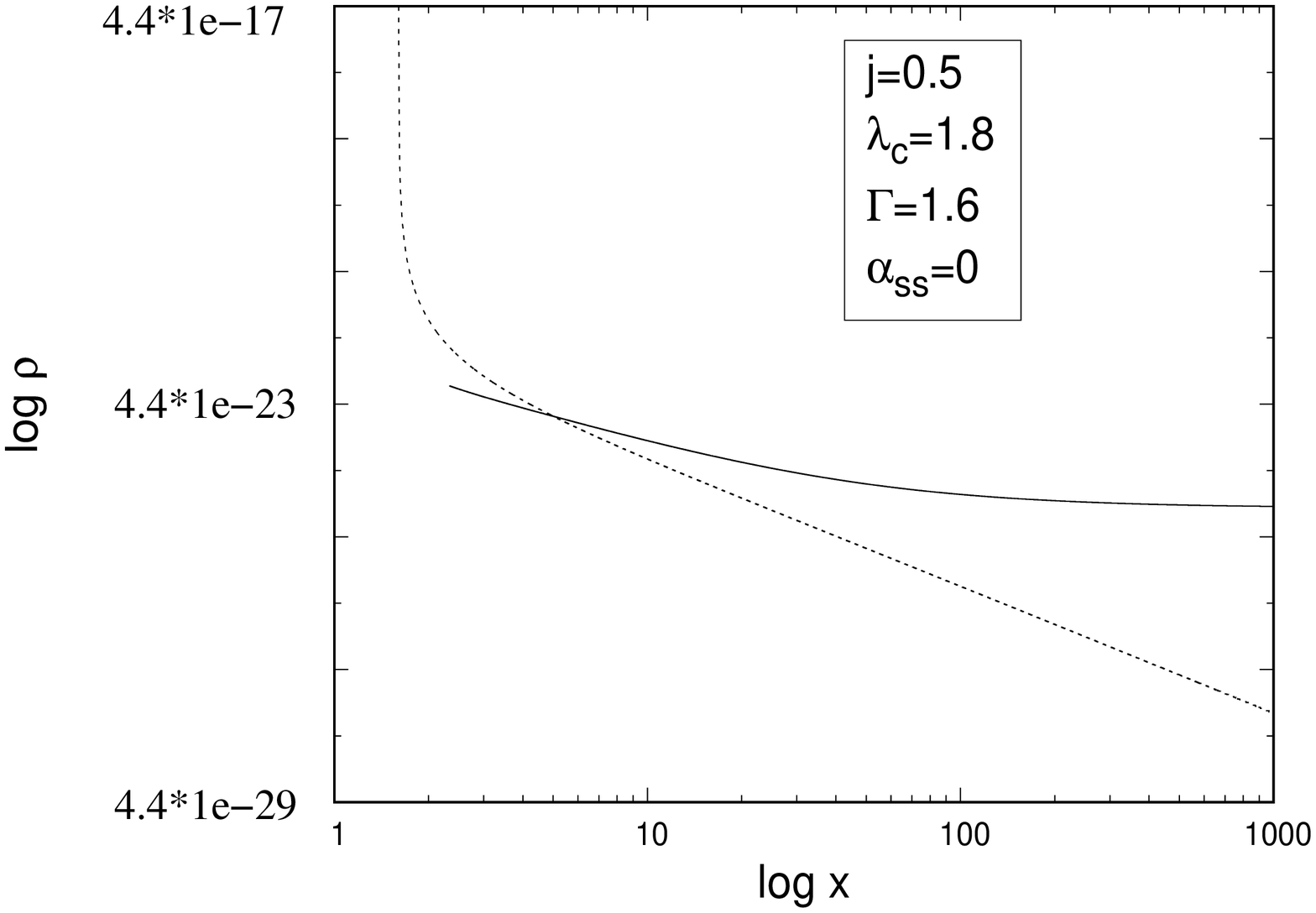}~~
\hspace*{0.2 in}
\includegraphics*[scale=0.3]{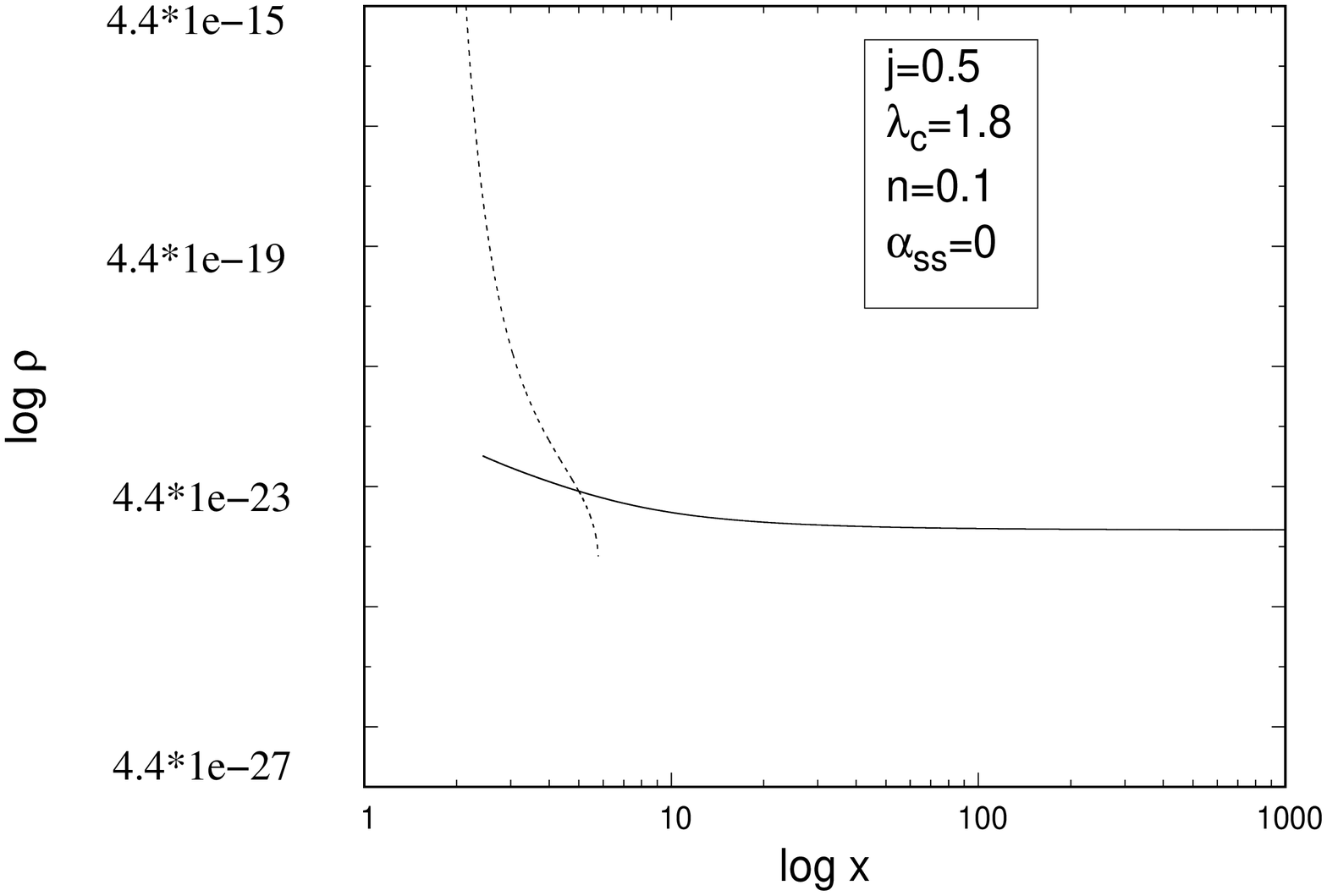}~~\\

~~~~~~~Fig $2.1.2.a$~~~~~~~\hspace{3 in}~~~~~~~~~Fig $2.1.2.b$
\includegraphics*[scale=0.3]{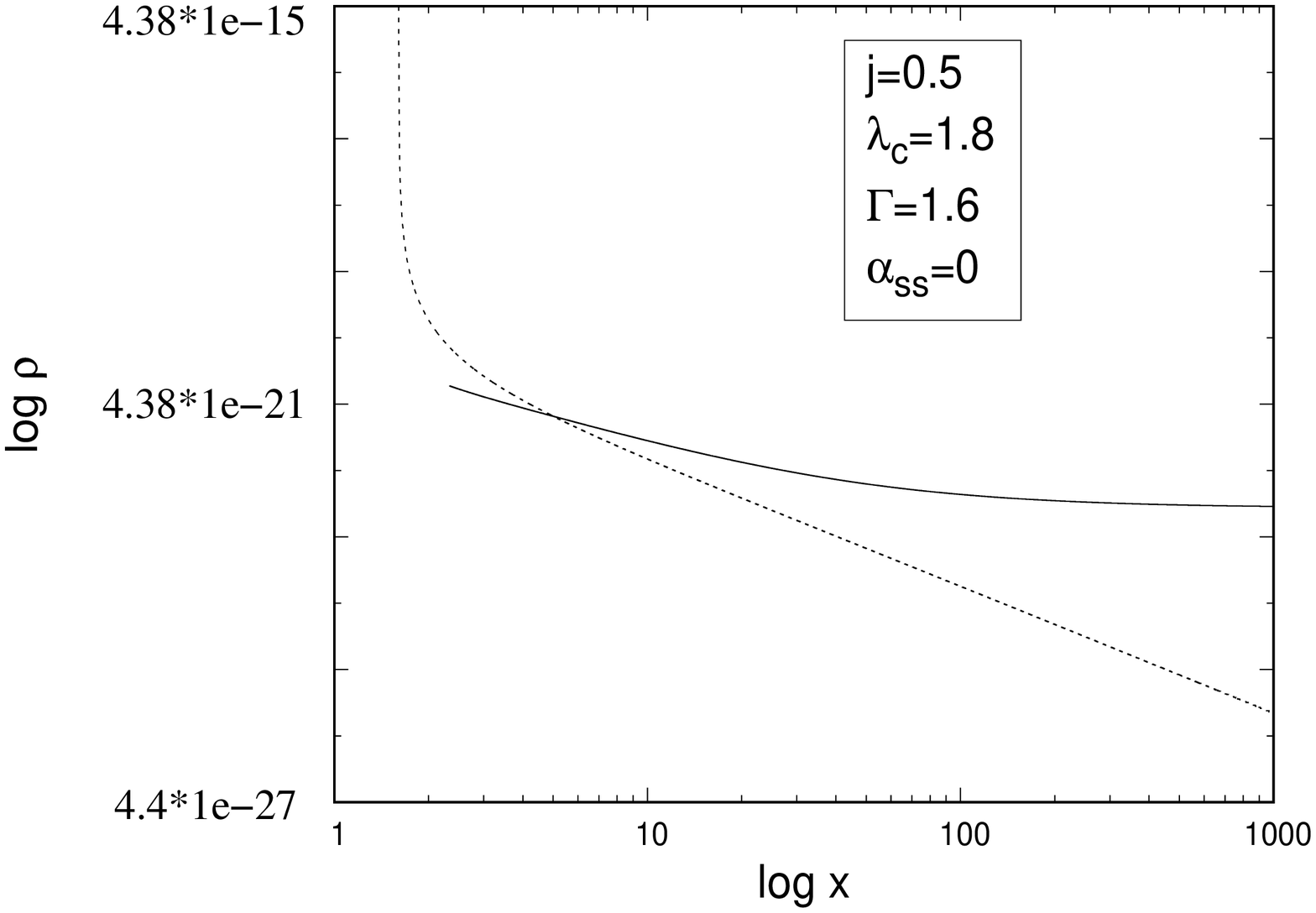}~~
\hspace*{0.2 in}
\includegraphics*[scale=0.3]{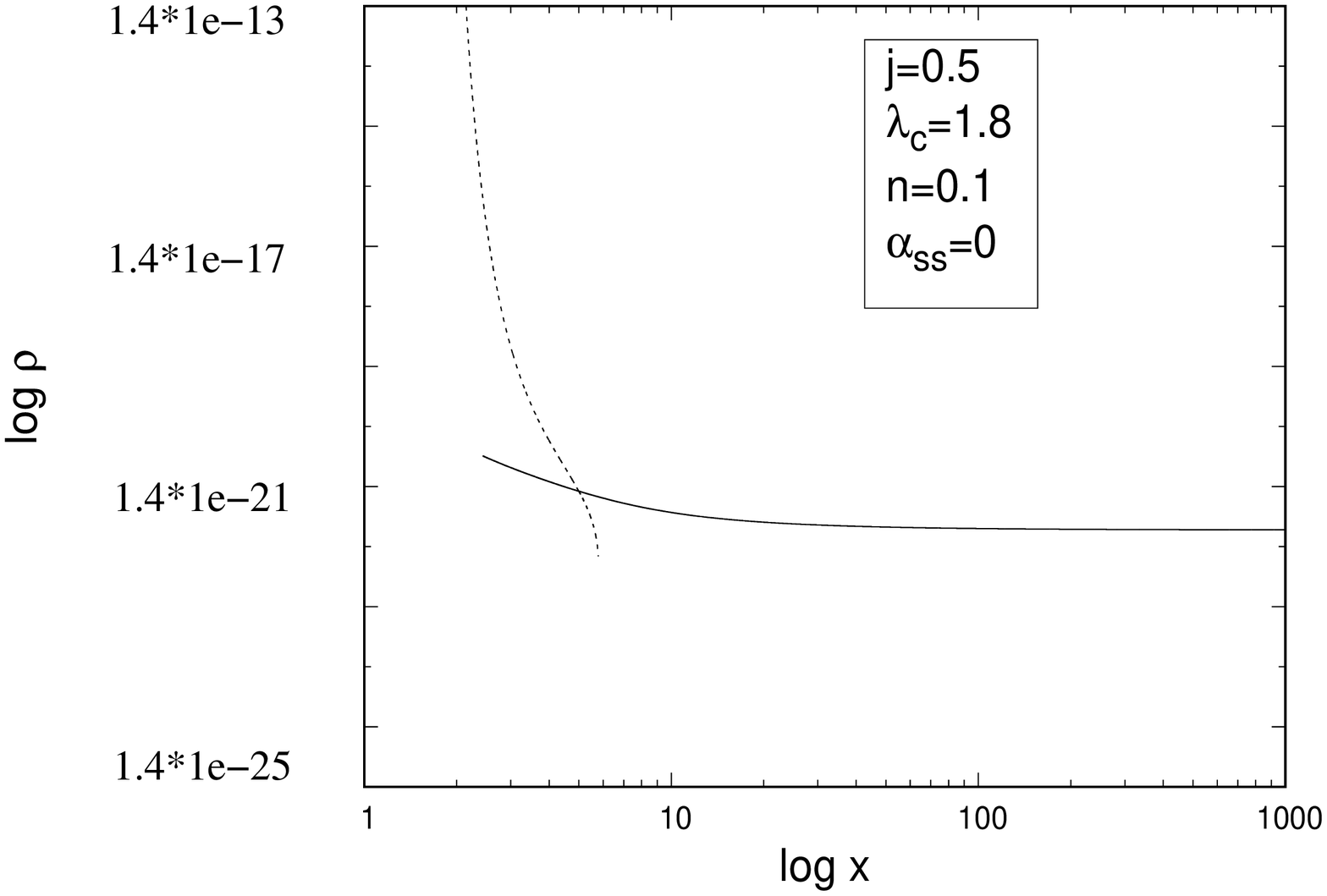}~~\\
\it{Figures $2.1.1.a$ and $2.1.2.a$ are plots of $log(\rho)$ vs $log(x)$ for nonviscous accretion disc flow around a non-rotating BH for adiabatic case using NFW and Einasto profile respevctively. Figures $2.1.1.b$ and $2.1.2.b$ are plots of $log(\rho)$ vs $log(x)$ for nonviscous accretion disc flow around a non-rotating BH for MCG using NFW and Einasto profile respectively. Accretion and wind curves are depicted by solid and dotted lines respectively.}
\end{figure}
\begin{figure}[h!]
\centering
~~~~~~~Fig $2.2.1.a$~~~~~~~\hspace{3 in}~~~~~~~~~Fig $2.2.1.b$
\includegraphics*[scale=0.3]{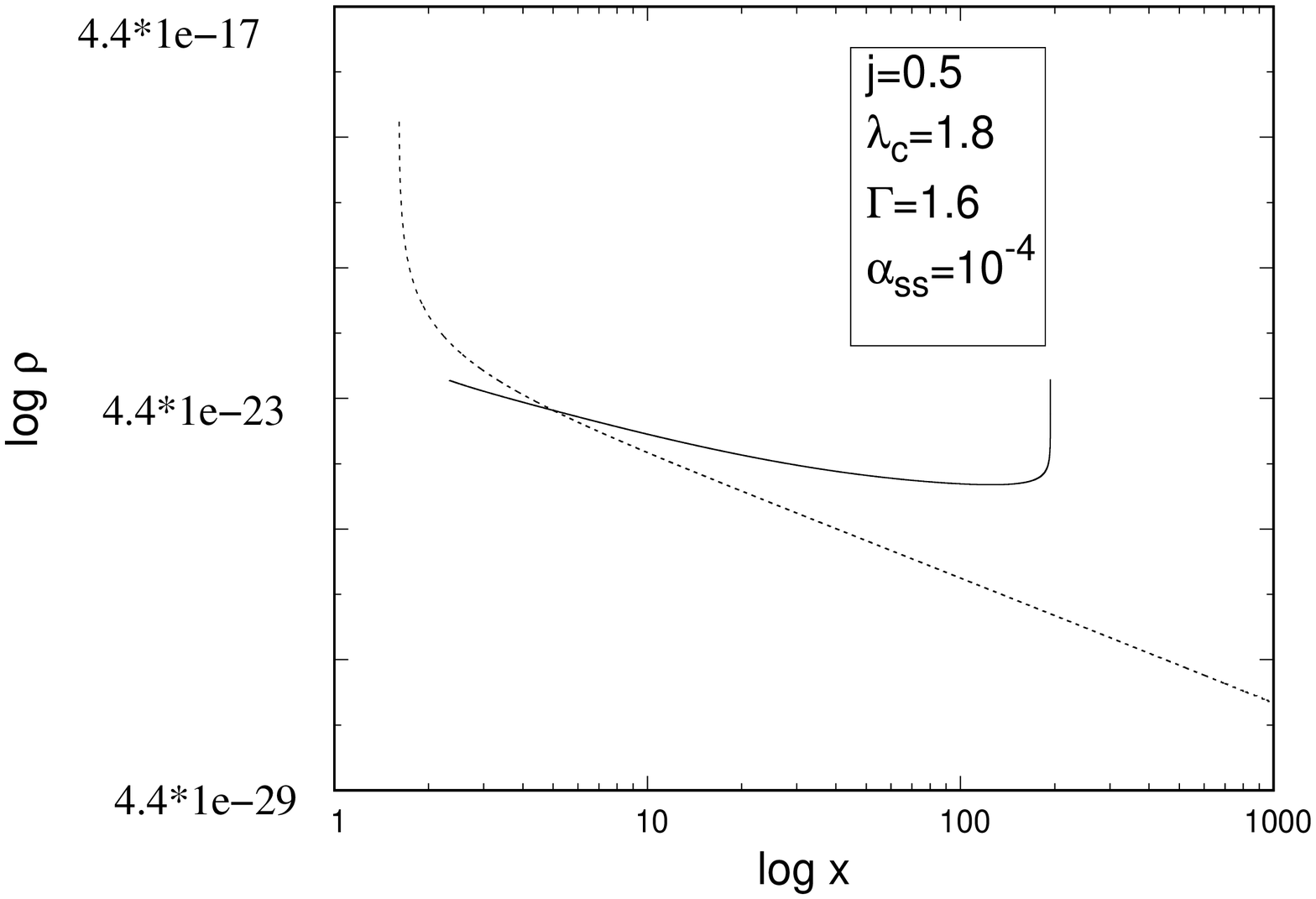}~~
\hspace*{0.2 in}
\includegraphics*[scale=0.3]{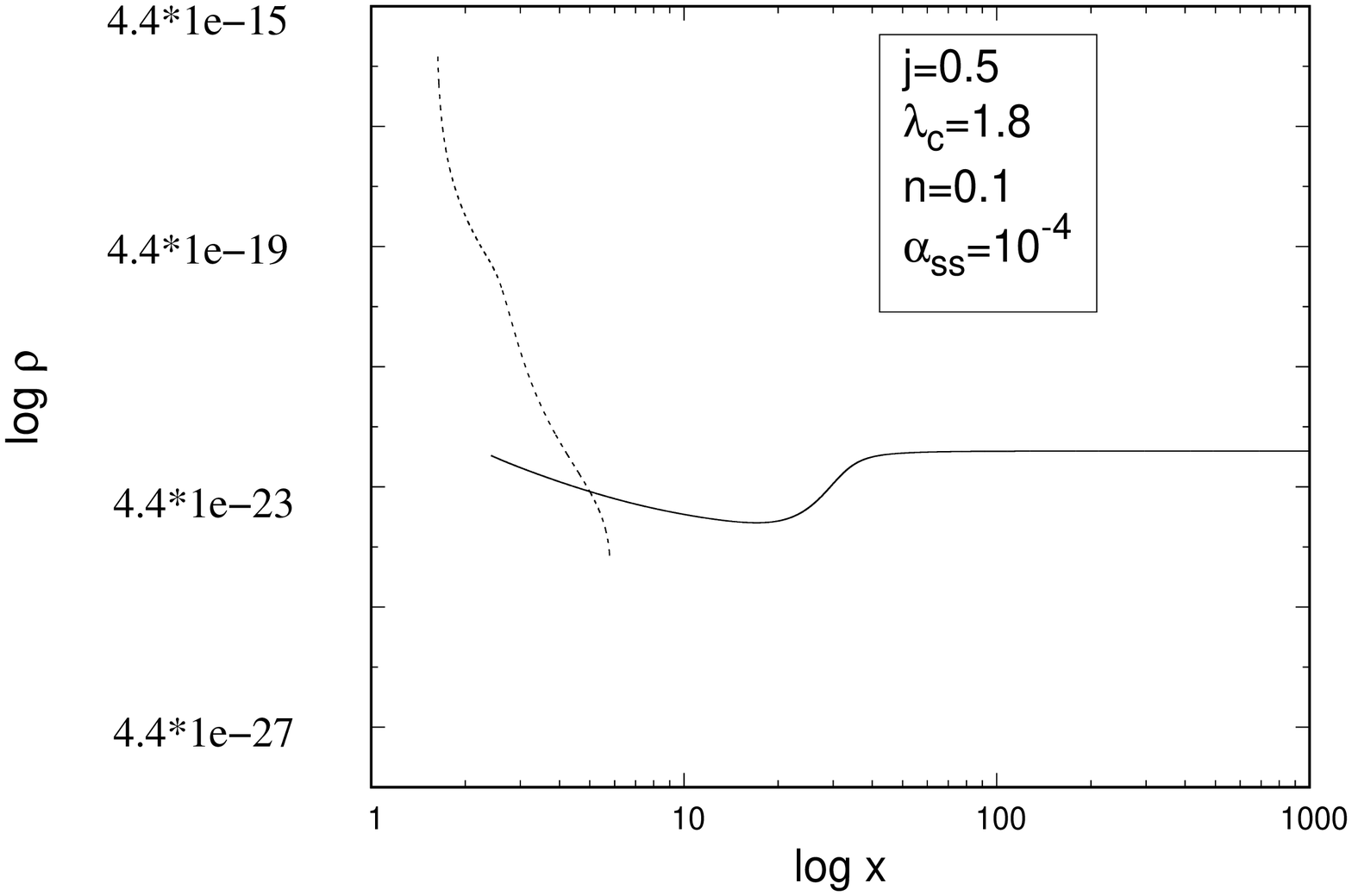}~~\\

~~~~~~~Fig $2.2.2.a$~~~~~~~\hspace{3 in}~~~~~~~~~Fig $2.2.2.b$
\includegraphics*[scale=0.3]{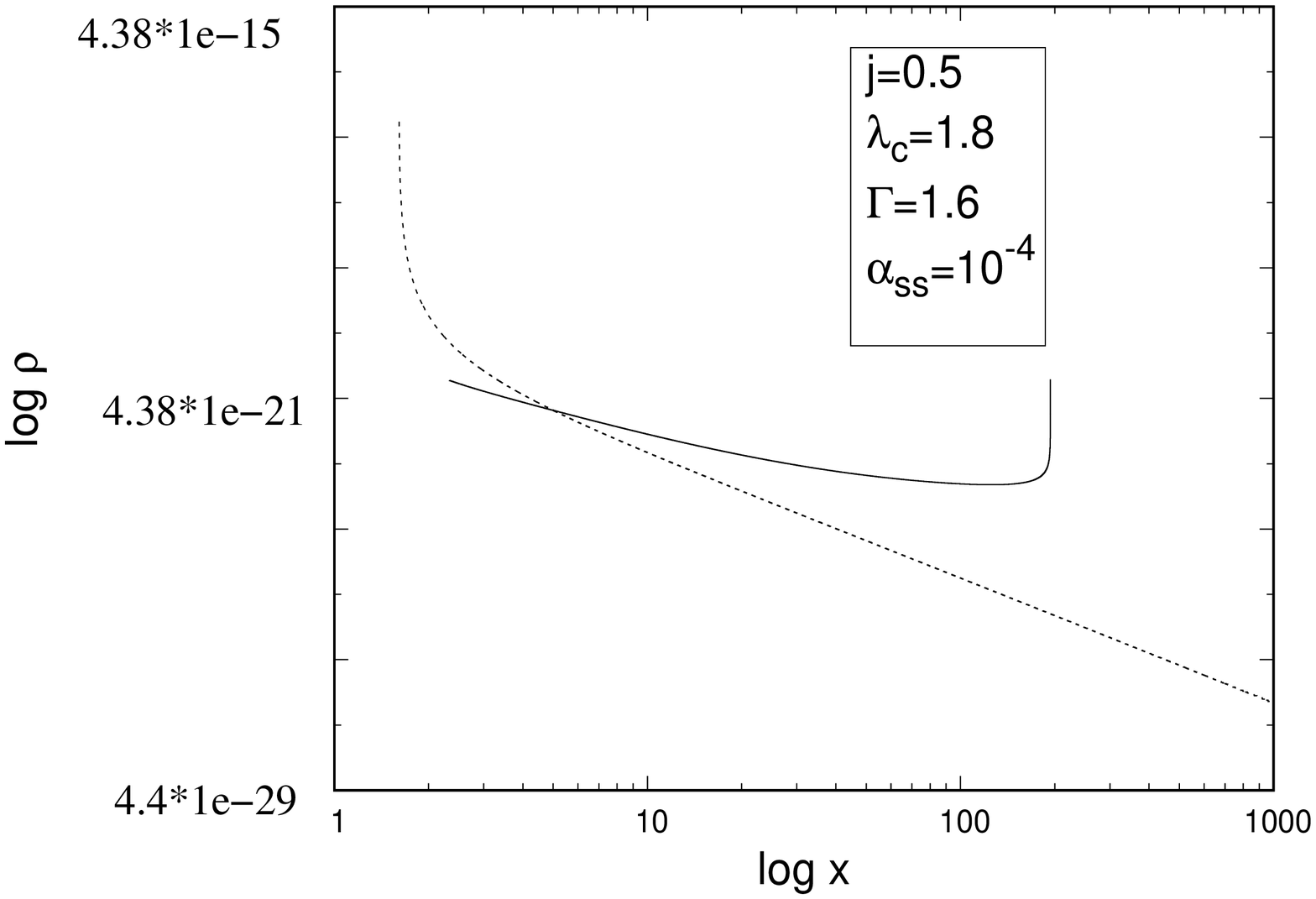}~~
\hspace*{0.2 in}
\includegraphics*[scale=0.3]{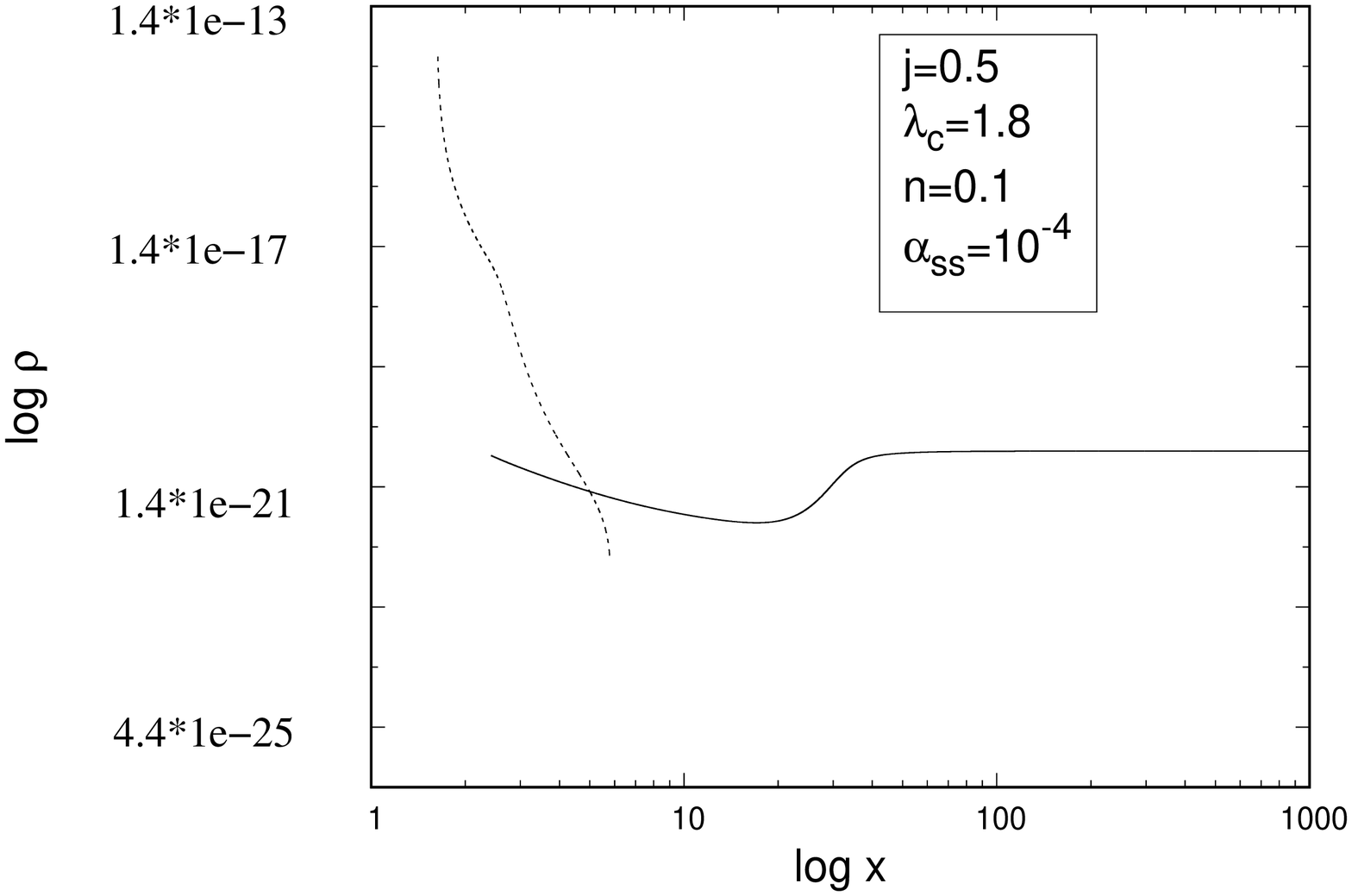}~~\\
\it{Figures $2.2.1.a$ and $2.2.2.a$ are plots of $log(\rho)$ vs $log(x)$ for nonviscous accretion disc flow around a non-rotating BH for adiabatic case using NFW and Einasto profile respevctively. Figures $2.2.1.b$ and $2.2.2.b$ are plots of $log(\rho)$ vs $log(x)$ for viscous accretion disc flow with viscosity $\alpha_{ss}=10^{-4}$, around a non-rotating BH for MCG using NFW and Einasto profile respectively. Accreti0n and wind curves are depicted by solid and dotted lines respectively.}
\end{figure}
\begin{figure}[h!]
\centering
~~~~~~~Fig $2.3.1.a$~~~~~~~\hspace{3 in}~~~~~~~~~Fig $2.3.1.b$
\includegraphics*[scale=0.3]{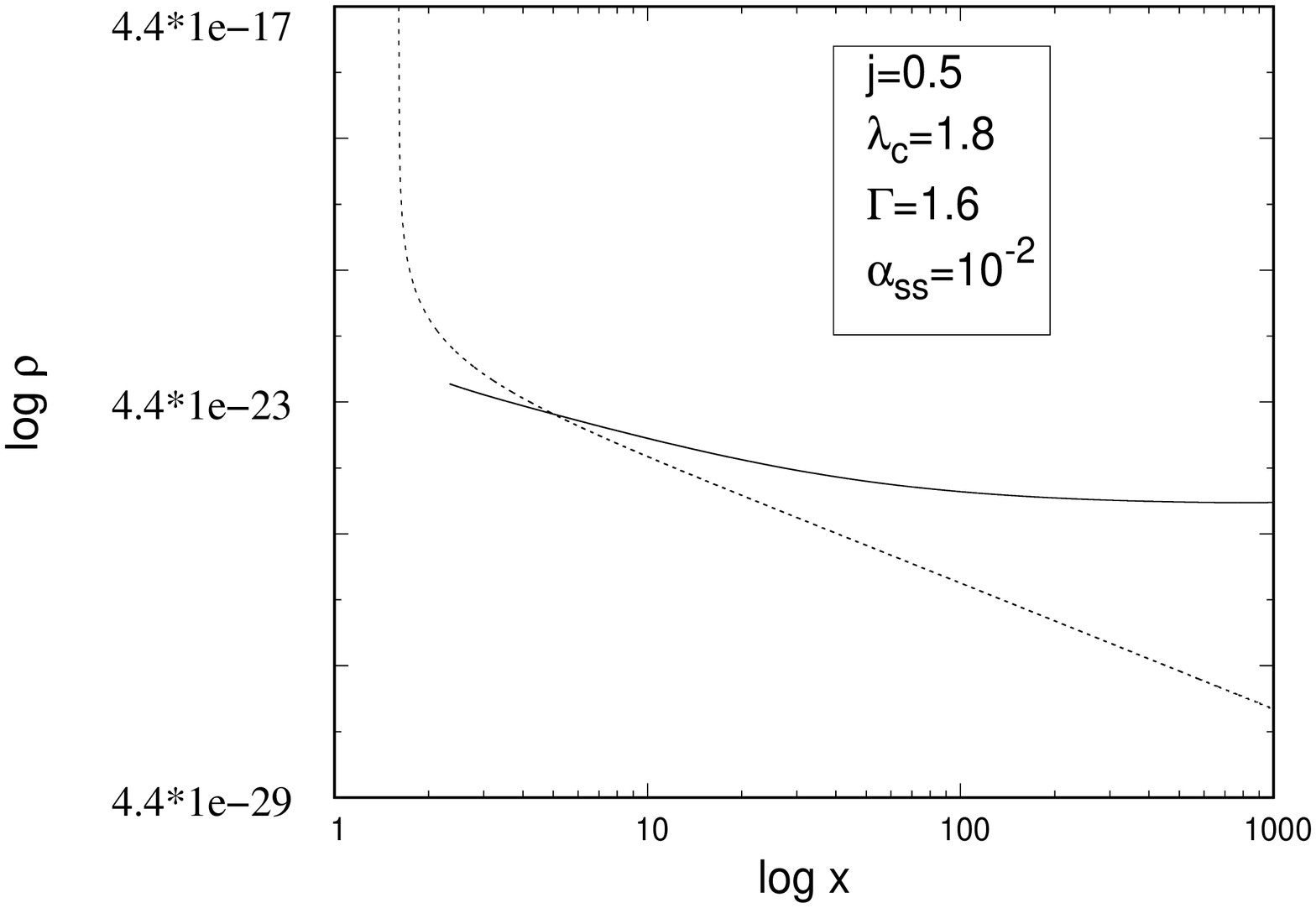}~~
\hspace*{0.2 in}
\includegraphics*[scale=0.3]{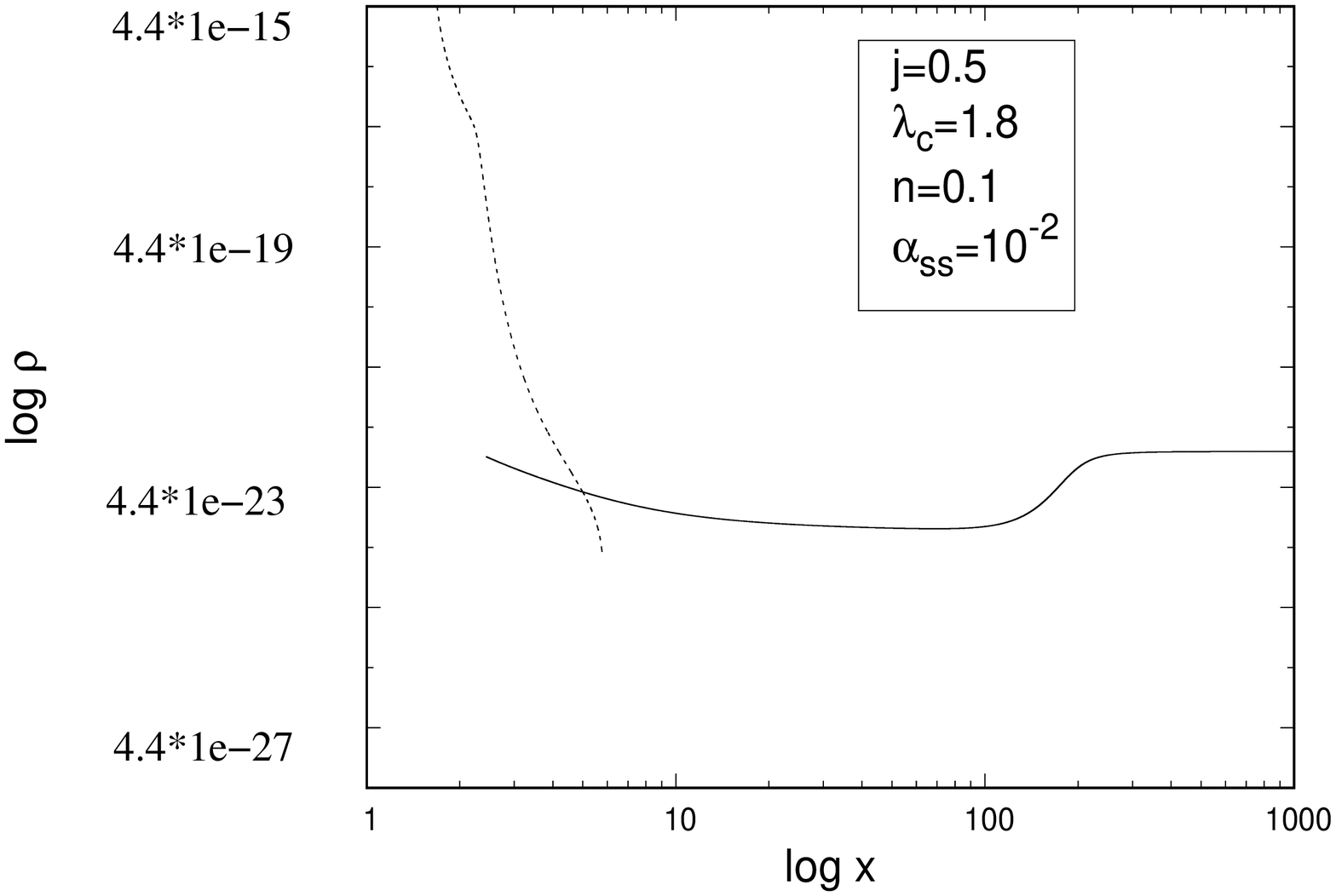}~~\\

~~~~~~~Fig $2.3.2.a$~~~~~~~\hspace{3 in}~~~~~~~~~Fig $2.3.2.b$
\includegraphics*[scale=0.3]{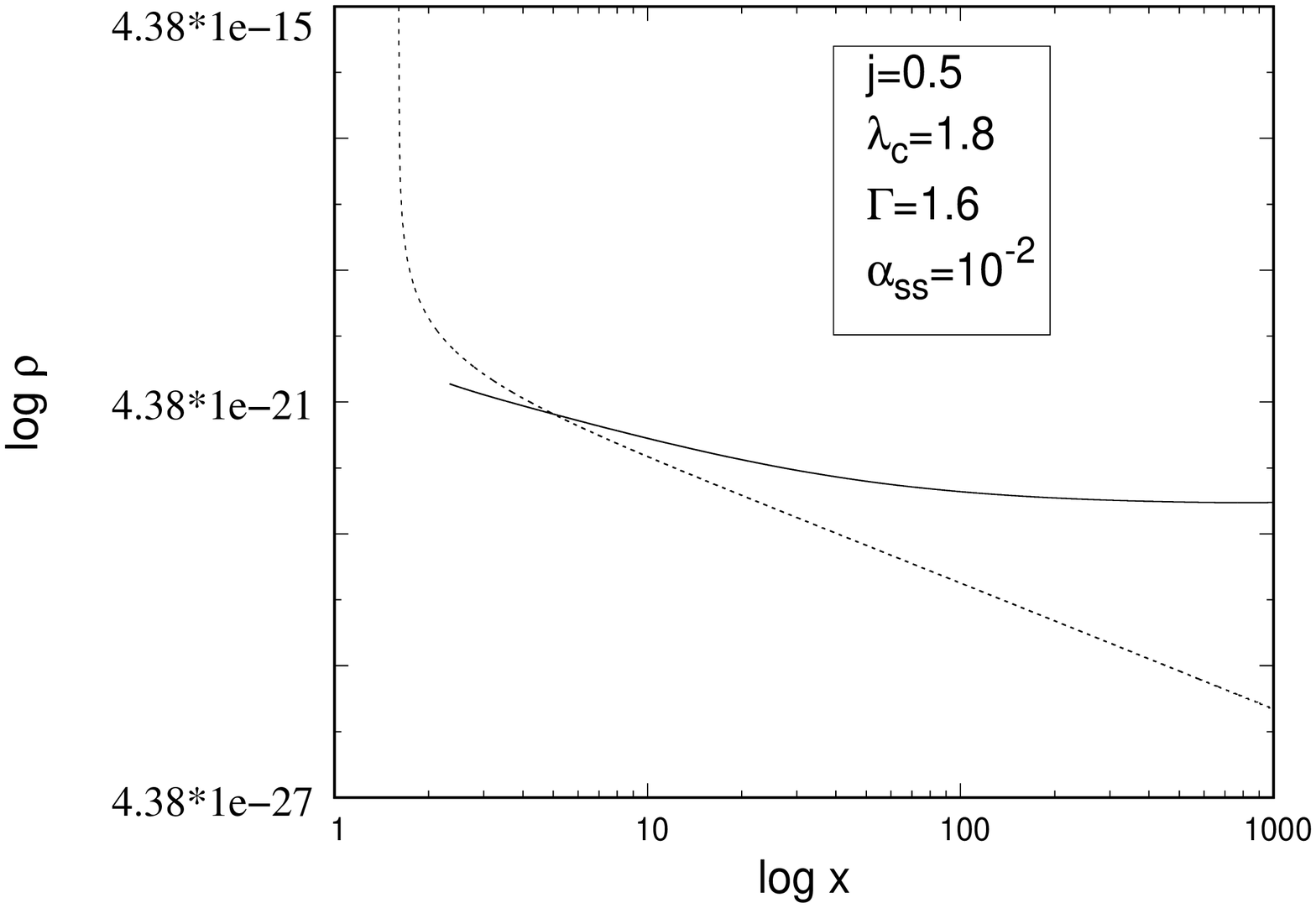}~~
\hspace*{0.2 in}
\includegraphics*[scale=0.3]{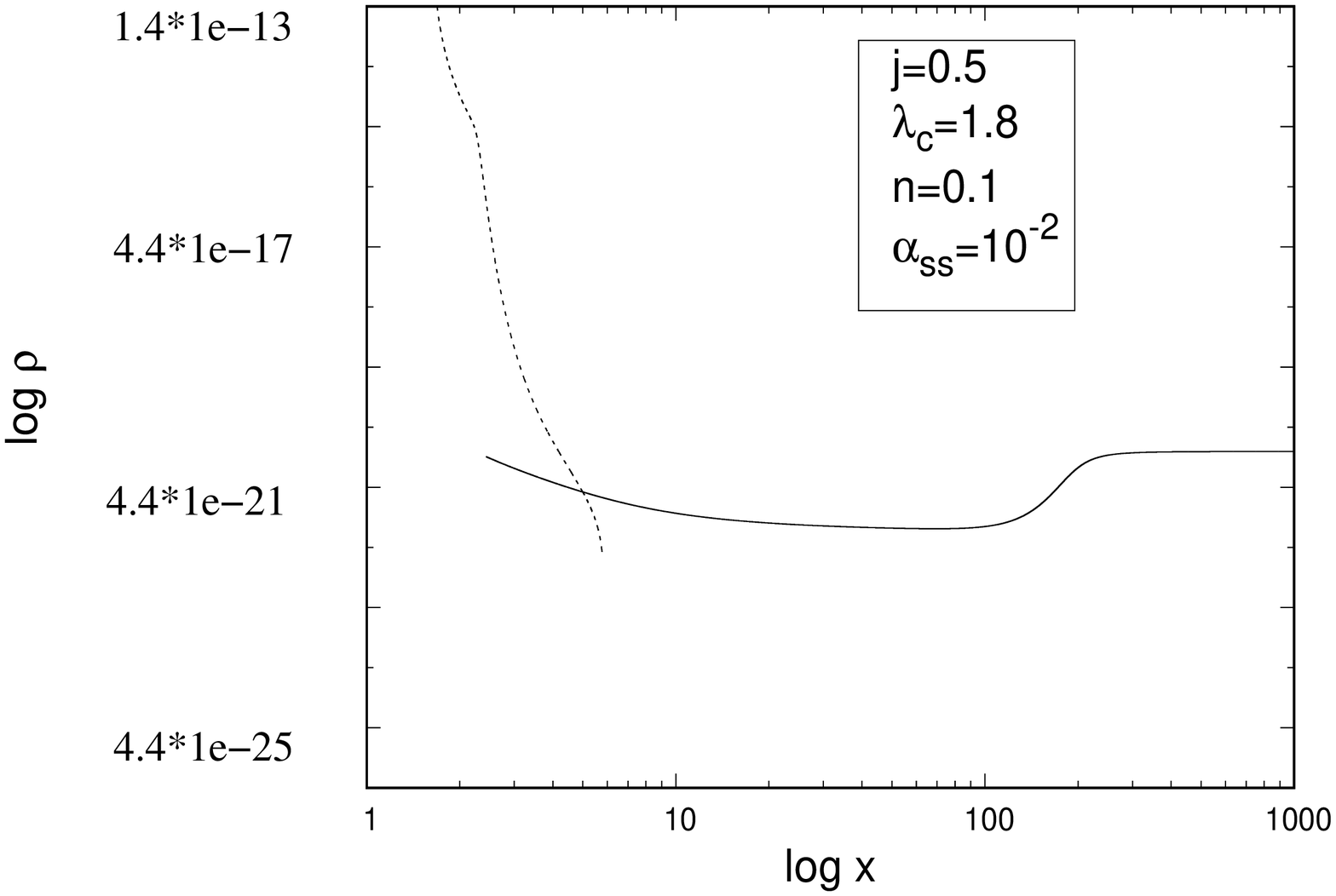}~~\\
\it{Figures $2.3.1.a$ and $2.3.2.a$ are plots of $log(\rho)$ vs $log(x)$ for nonviscous accretion disc flow around a non-rotating BH for adiabatic case using NFW and Einasto profile respevctively. Figures $2.3.1.b$ and $2.3.2.b$ are plots of $log(\rho)$ vs $log(x)$ for viscous accretion disc flow with viscosity $\alpha_{ss}=10^{-2}$, around a non-rotating BH for MCG using NFW and Einasto profile respectively. Accretion and wind curves are depicted by solid and dotted lines respectively.}
\end{figure}

In figure $2.1.1.a$ we have plotted density profile for non-viscous adiabatic accretion for a rotating BH with $j=0.5$. It is observed that for adiabatic accretion the wind density very near to BH is $10^4$ times lighter than the non-rotating case if the viscosity is not working. Even far from the BH, when we are thousand Schwarzchild radius distance apart from the BH, the wind density for adiabatic fluid is ten times lighter than the non-rotating case. However, the accretion density is more or less identical to the non-rotating case. 

In $2.1.1.b$ we have plotted the density for accreting Chaplygin gas onto a rotating BH with $j=0.5$. No viscosity is considered. Contrary to the adiabatic case we see that the wind density is hundred times heavier than the non-rotating case when we are very near to the BH. Rotation makes the disc radius shorter. This points out to the fact that dark energy's repulsive effect is inhibited by the rotation of BH. 

Fig $2.1.2.a$ and $2.1.2.b$ are the curves for Einasto profiles. The basic nature match with $2.1.1.a$ and $2.1.1.b$ respectively.

In fig $2.2.1.a$ to $2.2.2.b$ we have plotted the density profiles for rotating BH ($j=0.5$).
Viscosity started to act here ($\alpha_{ss}=10^{-4}$). The order of density in the unit $g/cm^3$ is equal to the non-viscous case. But for both the adiabatic and MCG case we see the wind branch ended very near to the BH (than the non-viscous rotating case). In addition, the accreting branch of the adiabatic case is formed to grow high as we go far from the BH (at a particular distance). For MCG if we go far from BH the accretion density increases abruptly at a certain point and then becomes constant. This signifies that the accreting MCG's density is constant at far. Then as it goes near to the BH the density falls at a particular distance (say ${\cal X}_{fall}$). Then it increases very slowly. So if viscosity and BH rotation both are working as a catalyst and an inhibitor, not only the wind is strengthened, but also accretion is been weakened at a certain distance.

In the figures $2.3.1.a$ to $2.3.2.b$ the viscosity is increased ($\alpha_{ss}=10^{-2}$), keeping the rotation same. For the MCG accretion case we see ${\cal X}_{fall}\vert_{\alpha_{ss}=10^{-2}} > {\cal X}_{fall}\vert_{\alpha_{ss}=10^{-4}}$. This means if viscosity is high, the weakening of accretion happens at further distance. This supports the conclusion that viscosity added with MCG decreases the accretion density as well.
\section{Brief Discussions and Conclusions}
In this letter, our motivation was to study the variations of density of accreting and wind fluids where DE is assumed to accrete onto SMBHs. Primarily, it is observed that for adiabatic flow if we are at a distance $x>x_{sonic}$ then $\rho_{wind}< \rho_{accretion}$. But once we are at the region $x< x_{sonic}$ then $\rho_{wind}> \rho_{accretion}$. Near to the BH, the wind density is very high. But if we consider DE to accrete in, the wind density increases remarkably. But the addition of viscosity gives us two distinct features. Firstly, the accretion disc's radius is more shortened with the viscous flow. The second important thing is the sudden reduction of accreting density. As we increase the viscosity, we observe that this sudden falling point goes far from the BH. This signifies if DE is treated to be the necessary parameter, more viscous is the flow, more shortened is the length from BH at which the wind velocity becomes equal to that of light. Besides this strong outward wind created by the repulsive nature of DE, we observe that the accretion density is also reduced at a distance. A threshold drop of density in accretion profile can be pointed out clearly. So DE not only increases the power of wind it also reduces the power of accretion if viscosity and rotation are working together with it. To match the results found by us is compared with figure 0. We observe that our results are staying in the range of density predicted by the reference \cite{K BoshkayevandD Malafarina}. 

\newpage

{\bf Achkowledgement : } This research is supported by the project grant of Government of West Bengal, Department of Higher Education, Science and Technology and Biotechnology (File no:- $ST/P/S\&T/16G-19/2017$). RB thanks IUCAA for providing Visiting Associateship. RB also thanks Prof. Banibrata Mukhopadhyay, Department of Physics, IISc, Bangalore, India-560012 for a detailed discussion regarding this problem earlier. SD thanks to Department of Higher Education, Government of West Bengal for research fellowship.


\begin{thebibliography}{100}
\bibitem{Eventhorizontelescope} The Event Horizon Telescope Collaboration; Akiyama, K. et al :- ``First M87 Event Horizon Telescope Results. I. The Shadow of the Supermassive Black Hole";  {\it The Astrophysical Journal Letters}, {\bf 875}, 1(2019);  astro-ph/1906.11238.
\bibitem{mail_gyan} Bambi, C. et al :-``Testing the rotational nature of the supermassive object M87* from the circularity and size of its first image", {\it Phys Rev D} (accepted); arXiv : arXiv:1904.12983 [gr-qc].
\bibitem{Gravitational wave} Abbott, B. P. et al. :-  ``Observation of Gravitational Waves from a Binary Black Hole Merger", {\it Phys. Rev. Lett.},{\bf 116} (2016)061102.
\bibitem{Eduardo Banados} Banados, E. et al :- ``An 800-million-solar-mass black hole in a significantly neutral Universe at a redshift of 7.5"; {\it Nature} {\bf 553}, 473(2018); 1712.01860v2 [astro-ph.GA].
\bibitem{K BoshkayevandD Malafarina} K Boshkayev, K. and Malafarina, D. :- ``A model for a dark matter core at the Galactic centre"; {\it Monthly Notices of the Royal Astronomical Society} {\bf 484} Issue 3(2019),  3325;  arXiv:1811.04061v2 [gr-qc].
\bibitem{PRLinteraction} M. S. Turner, G. Steigman, and L. M. Krauss, Phys. Rev.Lett.52, 2090 (1984).
\bibitem{PRDinteraction} Zimdahl, D. J. Schwarz, A. B. Balakin, and D. Pavon,Phys. Rev. D64, 063501 (2001)
\bibitem{Michel} Michel, F. C. :- `` Accretion of Matter by Condensed Objects",
{\it Astrophysics and Space Science}, {\bf 15} (1972)153-160.
\bibitem{Bondi} Bondi, H.:- ``On Spherically Symmetrical Accretion", {\it MNRAS}, {\bf 112} (1952)194-205.
\bibitem{ShakuraSunyaev1973} Shakura, N. I. , Sunyaev, R. A. :- ``Black Holes in Binary System. Observational Appearance", {\it  Astron. \& Astrophys.}, {\bf 24} (1973)337-355 .
\bibitem{NT} Novikov, I., and Thorne, K. S. 1973, in Black Holes, ed. C. DeWitt and B. S. Dewitt (New York: Gordon and Breach), 343
\bibitem{Paczynsky and Witta} Paczynsky, B. and Witta, P.J. :- {\it Astron. Astrophys.}, {\bf 88} (1980)23-31.
\bibitem{pseudonewtonian} Mukhopadhyay, B. :- ``Description of Pseudo-Newtonian Potential for the Relativistic Accretion Disks around Kerr Black Holes", {\it The Astrophysical Journal}, {\bf 581} (2002)1 [arXiv:astro-ph/0205475].
\bibitem{Mukhopadhyay2003} Mukhopadhyay, B. :- ``Stability of accretion disk around rotating black holes: a pseudo-general-relativistic fluid dynamical study" {\it Astrophys.J.}, {\bf  586} (2003) 1268-1279 [arXiv : astro-ph/0212186].
\bibitem{viscocitybmukherjee1} Mukhopadhyay, B. , Ghosh, S. :- ``Global solution of viscous accretion disk around rotating compact objects: A Pseudo-general-relativistic study", {\it Mon.Not.Roy.Astron.Soc.}, {\bf 342} (2003) 274-286 [arXiv: astro-ph/0304157].
\bibitem{Babichev2004 DE accretion PRL}  Babichev, E., Dokuchaev, V. ,  Eroshenko, Yu.:- ``Black hole mass decreasing due to phantom energy accretion", {\it Phys.Rev.Lett.}, {\bf  93} (2004) 021102. [arXiv :  gr-qc/0402089].
\bibitem{Peebles} Peebles, P. J. E., Ratra, B.:- ``The cosmological constant and dark energy". {\it Reviews of Modern Physics.}, {\bf 75} (2003) 559–606. [arXiv:astro-ph/0207347].
\bibitem{Bento} Bento, M. C. , Bertolami, O. , Sen, A. A. :- ``Generalized Chaplygin gas, accelerated expansion and dark energy matter unification", {\it Phys.Rev. D}, {\bf 66} (2002) 043507 [	arXiv: gr-qc/0202064].
\bibitem{Eos1} Debnath, U. et. al.:- ``Role of modified Chaplygin gas in accelerated universe", {\it Classical and Quantum Gravity}, {\bf 21} (2004) 23 [arXiv: gr-qc/0411015].
\bibitem{Eos2} Benaoum, H. B. :- ``Accelerated Universe from Modified Chaplygin Gas and Tachyonic Fluid", [ arXiv:hep-th/0205140].
\bibitem{Biswasaccretion1} Biswas, R. et. al. :-``Accretion of Chaplygin gas upon black holes: formation of faster outflowing winds", {\it Classical and Quantum Gravity}, {\bf 28} (2011) 035005 [arXiv: 1101.4602/ astro-ph].
\bibitem{Biswasaccretion2} Biswas, R. :-``Density profiles for Chaplygin gas accretion upon black holes: Moderately differentiated minima in wind branch", {\it Europhys.Lett.},{\bf 96} (2011) 49001.
\bibitem{Jamil1} Jamil, M. :- ``Evolution of a Schwarzschild black hole in phantom-like Chaplygin gas cosmologies" {\it Eur.Phys.J. C}, {\bf 62} (2009) 609-614 [arXiv:0906.2875/ gr-qc].
\bibitem{Sandip1} Dutta, S., Biswas, R. :- ``Fate of an Accretion Disc around a Black Hole when both the Viscosity and Dark Energy is Effecting" {\it Euro. Phys. J. C } {\bf 77 } (2017) 717 [arXiv:1705.11058/astro-ph.HE].
\bibitem{Deimer} Deimer, B. and and Kravtsov, A. V. :- ``Dependence of the outer density profiles of halos on their mass cccretion Rate", {The Astrophysical Journal}, {\bf 789} (2014)18 [arXiv: 1401.1216/ astro-ph].
\bibitem{Salvador-Sole} Salvador-Sole, E. et al :-``The nature of dark matter and the density profiles and central behaviour of relaxed halos", {\it The Astrophysical Journal}, {\bf 666} (2007)181-188. 

\bibitem{Navarro_2004} Navarro, J. F.et al.:- ``The inner structure of ΛCDM halos — III. Universality and asymptotic slopes", {\it Mon. Not. Roy. Astron. Soc.},{\bf 349} (2004) 1039 [arXiv: astro-ph/03111231].
\bibitem{Stadel_2009} Stadel, J. et al.:- ``Quantifying the heart of darkness with GHALO — a multi-billion particle simulation of our galactic halo", {\it Mon. Not. Roy. Astron. Soc.},{\bf 398} (2009) L21 [arXiv: 0808.2981/ astro-ph].
\bibitem{Navarro_1997} Navarro, F. J. et. al.:-``A universal density profile from hierarchical clustering", {\it The Astrophysical Journal}, {\bf 490}, (1997)493-508  [arXiv: astro-ph/9611107].
\bibitem{Diemand_2004} Diemand, J., Moore, B., and Stadel, J.:- ``Convergence and scatter of cluster density profiles", {\it Mon. Not. Roy. Astron. Soc.}, {\bf 353} (2004) 624  [arXiv: astro-ph/ 0402267].
\bibitem{Diemand_2005a} Diemand, J., Moore, B., and Stadel, J.:- ``Earth-mass dark-matter haloes as the first structures in the early universe", {\it Nature},{\bf 433} (2005) 389 [arXiv: astro-ph/0501589].
\bibitem{Diemand_2005b} Diemand, J. et al.:- ``Cusps in cold dark matter haloes",
{\it Mon. Not. Roy. Astron. Soc.}, {\bf 364} (2005) 665 [arXiv: astro-ph/0504215].
\bibitem{Diemand_2011} Diemand, J. and Moore, B.:- ``The structure and evolution of cold dark matter halos", {\it Adv. Sci. Lett.},{\bf 4} (2011) 297 [arXiv: 0906.4340/ astro-ph].
\bibitem{Ludlow_2011} Ludlow, A. D. et al.:- ``The density and pseudo-phase-space density profiles of cold dark matter haloes", {\it Mon. Not. Roy. Astron. Soc.},{\bf 415} (2011) 3895 [arXiv: 1102.0002/ astro-ph].
\bibitem{Navarro_2010} Navarro, J.F. et al.:- ``The diversity and similarity of cold dark matter halos", {\it Mon. Not. Roy. Astron. Soc.}, {\bf 402} (2010) 21 [arXiv: 0810.1522/ astro-ph].
\bibitem{Tasitsiomi_2004} Tasitsiomi, A. et al. :- ``Density profiles of ΛCDM clusters", {\it Astrophys. J.}, {\bf 607} (2004) 125 [arXiv: astro-ph/311062].
\bibitem{Flores_1994} Flores, R. A. and Primack, J. R.:- ``Observational and theoretical constraints on singular dark matter halos", {\it Astrophys. J.},{\bf 427} (1994) L1 [arXiv: astro-ph/9402004].
\bibitem{Dubinski_1991} Dubinski, J. and Carlberg, R. G.:- ``The structure of cold dark matter halos",{\it Astrophys. J.},{\bf  378} (1991) 496.
\bibitem{Pontzen_2014} Pontzen, A. and Governato, F. :- ``Cold dark matter heats up", {\it Nature}, {\bf 506} (2014) 171 [arXiv: 1402.1764/ astro-ph].
\bibitem{Weinberg_2015} Weinberg, D. H., Bullock, J. S., Governato, F., Kuzio de Naray, R. and Peter, A.H.G.:- ``Cold dark matter: controversies on small scales", {\it P. Natl. Acad. Sci.}, {\bf 112} (2015) 12249.
\bibitem{Remus_2013} Remus, R. S. et al.:- ``The dark halo — spheroid conspiracy and the origin of elliptical galaxies", {\it Astrophys. J.}, {\bf 766} (2013) 71  [arXiv: 1211.3420/ astro-ph].
\bibitem{Diemand_2008} Diemand, J. et al., ``Clumps and streams in the local dark matter distribution",{\it Nature}, {\bf 454} (2008) 735 [arXiv: 0805.1244/ astro-ph].
\bibitem{Merritt_2009} Merritt, D. :- ``Evolution of nuclear star clusters", {\it Astrophys. J.}, {\bf 694} (2009) 959 [arXiv: 0802.3186/ astro-ph].
\bibitem{Boker_2002} Böker, T. et al. :- ``A Hubble Space Telescope census of nuclear star clusters in late-type spiral galaxies. I. Observations and image analysis", {\it Astron. J.}, {\bf 123} (2002) 1389 [arXiv: astro-ph/ 0112086].
\bibitem{Newman_2013} Newman, A. B. et al. :- ``The density profiles of massive, relaxed galaxy clusters. I. The total density over three decades in radius", {\it Astrophys. J.}, {\bf 765} (2013) 24 [arXiv: 1209.1391/ astro-ph].
\bibitem{Chae_2014} Chae, K.H., Bernardi, M. and Kravtsov, A. V. :- `` Modelling mass distribution in elliptical galaxies: mass profiles and their correlation with velocity dispersion profiles", {\it Mon. Not. Roy. Astron. Soc.},{\bf 437} (2014) 3670 [arXiv: 1305.5471/ astro-ph].
\bibitem{Tal_2012} Tal, T., Wake, D.A. and van Dokkum, P.G. :- ``Observations of dark and luminous matter: the radial distribution of satellite galaxies around massive red galaxies", {\it Astrophys. J.}, {\bf 751} (2012) L5 [arXiv: 1201.5114/ astro-ph].
\bibitem{Williums_2010} Williams, L.L.R. and Hjorth, J. :- ``Statistical mechanics of collisionless orbits. II. Structure of halos", {\it Astrophys. J.}, {\bf 722} (2010) 856 [arXiv: 1010.0266/ astro-ph].
\bibitem{Hjorth_2010} Hjorth, J. and  Williams, L.L.R. :- `` Statistical mechanics of collisionless orbits. I. Origin of central cusps in dark-matter halos", {\it Astrophys. J.}, {\bf 722} (2010) 851 [arXiv: 1010.0265].
\bibitem{Feron_2008} Féron, C. and Hjorth, J. :- ``Simulated dark-matter halos as a test of nonextensive statistical mechanics", {\it Phys. Rev. E}, {\bf 77} (2008) 022106 [arXiv: 0801.2504].
\bibitem{Medvedev_2001} Medvedev, M. V. and Rybicki, G. :- ``The structure of selfgravitating polytropic systems with n around 5", {\it Astrophys. J.}, {\bf 555} (2001) 863 [arXiv: astro-ph/0010621].
\bibitem{Binney_1987} Binney, J. and Tremaine, S. :- ``Galactic dynamics", {\bf 1st ed.}, {\it Princeton University Press}, Princeton U.S.A. (1987).
\bibitem{King_1966} King, Ivan, R. :- ``The structure of star clusters. III. Some simple dynamical models", {\it Astro. J.}, {\bf 71} (1966) 64
\bibitem{Einasto_1965} Einasto, J. Trudy Astrofizicheskogo Instituta Alma-Ata, {\bf 5},(1965)87.

\bibitem{young_2016} Young, A. M. et. al.:- ``Ubiquity of density slope oscillations in the central regions of galaxy and cluster-sized systems", {\it JCAP}, {\bf 05} (2016) 010 [arXiv: 1604.01409v1/astro-ph].


\end{thebibliography}
\end{document}